\pdfoutput=1
\documentclass[final,5p,times]{elsarticle}
\usepackage[T1]{fontenc}
\usepackage[utf8]{inputenc}
\usepackage[final,activate={true,nocompatibility},kerning=true,spacing=none]{microtype}
\usepackage{hyperref}
\usepackage{listings}
\usepackage{siunitx}
\usepackage{graphicx}
\usepackage[table]{xcolor}
\usepackage{booktabs}
\usepackage{mathptmx}
\usepackage{xspace}
\usepackage{algorithm}
\usepackage[noend]{algpseudocode}
\usepackage{ntheorem}
\usepackage{amsmath}
\usepackage{amssymb}
\usepackage{paralist}
\usepackage{cleveref}
\usepackage{subcaption}

\algrenewcommand\algorithmicindent{0.3cm}

\crefrangelabelformat{line}{#3#1#4--#5#2#6}

\definecolor{kw}{rgb}{0.5,0.,0.33}
\definecolor{com}{rgb}{0.247,0.5,0.372}
\definecolor{str}{rgb}{0.165,0.0,1.0}
\definecolor{ano}{rgb}{0.4,0.4,0.4}
\definecolor{bg}{rgb}{1,1,1}

\lstdefinelanguage{MyJava}{
  stringstyle=\color{str}, 
  keywordstyle={\bfseries\color{kw}\ttfamily}, 
  commentstyle=\color{com}, 
  basicstyle={\scriptsize\ttfamily}, 
  captionpos=b,  
  showstringspaces=false,
  frame=single,
  escapechar=|, 
  numbersep=5pt, 
  numbers=left, 
  language=java, 
  tabsize=2,
  backgroundcolor=\color{bg}, 
  moredelim=[il][\textcolor{ano}]{$$},
  moredelim=[is][\textcolor{ano}]{\%\%}{\%\%}
}

\DeclareGraphicsExtensions{.pdf, .png, .jpg}
\graphicspath{{res/}}

\makeatletter
\newcommand\footnoteref[1]{\protected@xdef\@thefnmark{\ref{#1}}\@footnotemark}
\makeatother

\newcommand{\tool}{\texttt{InspectorGuidget}\xspace}
\newcommand{\ie}{\emph{i.e.,}\xspace}
\newcommand{\eg}{\emph{e.g.,}\xspace}
\newcommand{\cf}{\emph{cf.}\xspace}
\newcommand{\etal}{\emph{et al.}\xspace}

\newcommand{\bl}{\emph{Blob listener}\xspace}
\newcommand{\bls}{\emph{Blob listeners}\xspace}

\newtheorem{definition}{Definition}

\def\pprw{8.5in}
\def\pprh{11in}

\lstnewenvironment{code}[1][]%
  {\noindent\minipage{\linewidth}\medskip 
   \lstset{#1}}
  {\endminipage}
  
\definecolor{linkColor}{RGB}{6,125,233}

\journal{Information and Software Technology}

\begin{document}

\begin{frontmatter}
\title{User Interface Design Smell: Automatic Detection and Refactoring of Blob Listeners}

\author[AB]{Arnaud Blouin}
\ead{ablouin@irisa.fr}
\author[NM]{Valéria Lelli}
\ead{valerialelli@great.ufc.br}
\author[BB]{Benoit Baudry}
\ead{baudry@kth.se}
\author[FB]{Fabien Coulon}
\ead{fabien.coulon@irit.fr}
\address[AB]{Univ Rennes, INSA Rennes, Inria, CNRS, IRISA, France}
\address[NM]{Federal University of Ceará, Brazil}
\address[BB]{KTH Royal Institute of Technology, Sweden}
\address[FB]{University of Toulouse - Jean-Jaurès, France}

\begin{abstract}
\textbf{Context.}~User Interfaces (UIs) intensively rely on event-driven programming: 
interactive objects send UI events, which capture users' interactions, to dedicated objects called \emph{controllers}.
Controllers use several \emph{UI listeners} that handle these events to produce UI commands.\\
\textbf{Objective.}~First, we reveal the presence of design smells in the code that describes and controls UIs. 
Second, we demonstrate that specific code analyses are necessary to analyze and refactor UI code, because of its coupling with the rest of the code.\\
\textbf{Method.}~We conducted an empirical study on four large Java software systems.
We studied to what extent the number of UI commands that a UI listener can produce has an impact on the change- and fault-proneness of the UI listener code.
We developped a static code analysis for detecting UI commands in the code.\\
\textbf{Results.}~We identified a new type of design smell, called \bl, that characterizes UI listeners that can produce more than two UI commands.
We proposed a systematic static code analysis procedure that searches for \bl that we implement in \tool. 
We conducted experiments on the four software systems for which we manually identified 53 instances of \bl.
\tool successfully detected 52 \bls out of 53.
The results exhibit a precision of \SI{81.25}{\percent} 
and a recall of \SI{98.11}{\percent}.
We then developed a semi-automatically and behavior-preserving refactoring process to remove \bls.
\SI{49.06}{\percent} of the 53 \bls were automatically refactored.
Patches have been accepted and merged.
Discussions with developers of the four software systems assess the relevance of the \bl.\\
\textbf{Conclusion.}~This work shows that UI code also suffers from design smells that have to be identified and characterized.
We argue that studies have to be conducted to find other UI design smells and tools that analyze UI code must be developed.
\end{abstract}


\end{frontmatter}

\section{Introduction}\label{sec.intro}

User Interfaces (UI) are the tangible vector that enable users to interact with software systems.
While UI design and qualitative assessment is handled by UI designers, integrating UIs into software systems remains a software engineering task.
Software engineers develop UIs following widespread architectural design patterns, such as MVC~\cite{Krasner88} or MVP~\cite{potel1996} (\emph{Model}-\emph{View}-\emph{Controller}/\emph{Presenter}), that consider UIs as first-class concerns (\eg the \emph{View} in these two patterns).
These patterns clarify the implementations of UIs by clearly separating concerns, thus minimizing the "spaghetti of call-backs"~\cite{Myers91}. 
These implementations rely on event-driven programming where events are treated by \emph{controllers} (resp. presenters\footnote{For simplicity, we use the term \emph{controller} to refer to any kind of component of MV* architectures that manages events triggered by UIs, such as \emph{Presenter} (MVP), or \emph{ViewModel} (MVVM~\cite{smith09}).}), as depicted by \Cref{lst.introEX}.
In this Java Swing code example, the \emph{AController} controller manages three interactive objects, \emph{b1}, \emph{b2}, and \emph{m3} (\Crefrange{codeIntro1}{codeIntro11}).
To handle events that these objects trigger in response to users' interactions, the UI listener $ActionListener$ is implemented in the controller (\Crefrange{codeIntro3}{codeIntro4}).
One major task of UI listeners is the production of UI commands, \ie a set of statements executed in reaction to a UI event produced by an interactive object (\Cref{codeIntro5,codeIntro6,codeIntro10}).
Like any code artifact, UI controllers must be tested, maintained and are prone to evolution and errors.
In particular, software developers are free to develop UI listeners that can produce a single or multiple UI commands.
In this work, we investigate the effects of developing UI listeners that can produce one or several UI commands on the code quality of these listeners.

\begin{lstlisting}[language=MyJava, label=lst.introEX, caption={Code example of a UI controller}]
class AController implements ActionListener {
  JButton b1;|\label{codeIntro1}|
  JButton b2;|\label{codeIntro2}|
  JMenuItem m3;|\label{codeIntro11}|
  %%@Override%% public void actionPerformed(ActionEvent e) {|\label{codeIntro3}|
     Object src = e.getSource();
     if(src==b1){|\label{codeIntro7}|
        // Command 1|\label{codeIntro5}|
     }else if(src==b2)|\label{codeIntro8}|
        // Command 2|\label{codeIntro6}|
     }else if(src instanceof AbstractButton && 
         ((AbstractButton)src).getActionCommand().equals(|\label{codeIntro9}|
         m3.getActionCommand()))
        // Command 3|\label{codeIntro10}|
     }
}}|\label{codeIntro4}|
\end{lstlisting}

In many cases UI code is intertwined with the rest of the code.
The first step of our work thus consists of a static code analysis procedure that detects the UI commands that a UI listener can produce.
Using this code analysis procedure, we then conduct an empirical study on four large Java Swing and SWT open-source UIs:
\emph{Eclipse}, \emph{JabRef}, \emph{ArgouML}, and \emph{FreeCol}.
We empirically study to what extent the number of UI commands that a UI listener can produce has an impact on the change- or fault-proneness of the UI listener code,
considered in the literature as negative impacts of a design smell on the code~\cite{palomba14,Khomh2012,lozano2007assessing,rapu2004using}.
The results of this empirical study show evidences that UI listeners that control more than two commands are more error-prone than the other UI listeners.
Based on these results, we define a UI design smell we call \bl, \ie  a UI listener that can produce more than two UI commands.
For example with \Cref{lst.introEX}, the UI listener implemented in \emph{AController} manages events produced by three interactive objects, \emph{b1}, \emph{b2}, and \emph{m3} (\Cref{codeIntro7,codeIntro8,codeIntro9}), that produce one UI command each.
The empirical and quantitative characterization of the \bl completes the recent qualitative study on web developers that spots web controllers that do to much as a bad coding practice~\cite{aniche2017}. 

Based on the coding practices of the UI listeners that can produce less than three commands, we propose an automatic refactoring process to remove \bls.

We provide an open-source tool, \tool\footnote{\label{foot.webpage}\scriptsize\url{https://github.com/diverse-project/InspectorGuidget}}, that automatically detect and refactor \bls in Java Swing, SWT, and JavaFX UIs.
To evaluate the ability of \tool at detecting \bls, we considered the four Java software systems previously mentioned.
We manually retrieved all instances of \bl in each software, to build a ground truth for our experiments: 
we found 53 \bls. 
\tool detected 51 \bls out of 53.
The experiments show that our algorithm has a precision of \SI{88.25}{\percent} and recall of \SI{98.11}{\percent} to detect \bls.

We use the same four software systems to evaluate the ability of \tool at refactoring \bls.
\tool is able to automatically refactor \SI{49.06}{\percent} of the 53 \bls.
We show that two types of \bls exist and \tool is able to refactor \SI{86.7}{\percent} of one type of \bls.
Limitations of the UI toolkits limit the refactoring of the second type of \bls.
For the four software systems we submitted patches that remove the \bls and asked developers for feedback.
The patches for \emph{JabRef} and \emph{Freecol} have been accepted and merged.
The patches for \emph{Eclipse} are in review.
We received no feedback from \emph{ArgoUML} developers.
The concept of \bl and the refactoring solution we propose is accepted by the developers we interviewed.

Our contributions are:
\begin{enumerate}\itemsep-0.07cm
   \item an empirical study on four Java Swing and SWT open-source software systems. 
  This study investigates the current coding practices of UI listeners.
  The main result of this study is the identification of a UI design smell we called \bl.
   \item a precise characterization of the \bl.
We also discuss the different coding practices of UI listeners we observed in listeners having less than three commands.
   \item A static code analysis to automatically detect the presence of \bls in Java Swing, SWT, and JavaFX UIs.
   \item A code refactoring solution to remove \bl instances.
   \item an open-source tool, \tool, that embeds the code analysis and the code refactoring technique.
   \item A quantitative evaluation of the \bl detection and refactoring techniques.
   \item A qualitative evaluation of the \bl design smell and its refactoring solution.
   Patches were produced and submitted to the analyzed projects. 
   Projects from which we got answers have accepted and merged the submitted patches.
   Discussions with concerned developers were conducted on the relevance of: the Blob Listener design smell; the spotted instances of Blob Listener in their code; the refactoring solution.
\end{enumerate}

This paper extends our work published at EICS~2016~\cite{LEL16} with:
the refactoring solution, its implementation, and its evaluation;
a new algorithm and its implementation for detecting UI command, as the one proposed in~\cite{LEL16} contained errors and limitations;
a replication of the empirical study using this new implementation and on an improved data set composed of more representative software systems than the ones in~\cite{LEL16}.

The paper is organized as follows. 
\Cref{sec.cmd} introduces the concept of UI commands and the algorithm for automatically detecting UI commands in UI listeners.
Based on the implementation of this algorithm, called \tool, \Cref{sec.study} describes an empirical study that investigates coding practices of UI listeners.
This study exhibits the existence of an original UI design smell we called \bl. 
\Cref{sec.refactor} describes the refactoring solution for automatically removing \bls.
\Cref{sec.eval} evaluates the ability of \tool in detecting both UI commands and \bls, and in refactoring \bls.
The paper ends with related work (\Cref{sec.related}) and a research agenda (\Cref{sec.conclu}).

\section{Automatic Detection of UI Commands}\label{sec.cmd}

As any code artifact, UI code has to be validated to find bugs or design smells.
Design smells are symptoms of poor software design and implementation choices that affect the quality and the reliability of software systems~\cite{fowler1999}.
If software validation techniques are numerous and broadly used in the industry, they focus on specific programming issues, such as object-oriented issues. 
We claim that it is necessary to develop specific UI code analysis techniques to take the specific nature of UI code smells and bugs into account. 
In particular, these analysis techniques must embed rules to extract information specifically about UI code, while this one is deeply intertwined with the rest of the code.
These techniques have to extract from the code information and metrics related to UIs intertwined with the rest of the code.
In this section, we introduce a code analysis technique for detecting UI commands in the code of Java software systems.

\subsection{Definitions}

In this section we define and illustrate core concepts used in this work.

\begin{definition}[UI listener]
UI listeners are objects that follow the event-driven programming paradigm by receiving and treating UI events produced by users while interacting with UIs.
In reaction of such events, UI listeners produce UI commands.
UI toolkits provide predefined sets of listener interfaces or classes that developers can used to listen specific UI events such as clicks or key pressures.
\end{definition}

This definition follows the concept of \emph{Abstract Listener} from the W3C \emph{Abstract User Interface Models} report that defines a listener as the "\emph{entity used to describe the behavior of [an interactive object] by using Event-Condition-Action (ECA) rules}"~\cite{W3CMBUI}.
ECA in the context of UI listeners refers to UI events (\emph{Event}), conditional statements that guard the creation of UI commands (\emph{Condition}), and UI commands (\emph{Action}).
UI listeners are not limited to graphical user interfaces.
UI listeners aim at processing events produced by a user interface, which can be graphical, vocal, command line, \emph{etc.}.

\noindent UI commands can be defined as follows:

\begin{definition}[UI Command]
A UI command~\cite{GAM95,BEA00b}, aka. \emph{action}~\cite{BLO10,BLO11}, is a set of statements executed in reaction to a user interaction, captured by an input event, performed on a UI. 
UI commands may be supplemented with pre-conditions checking whether the command fulfills the prerequisites to be executed.
\end{definition}

\begin{figure}[h]
	\centering
	\includegraphics[width=0.85\columnwidth]{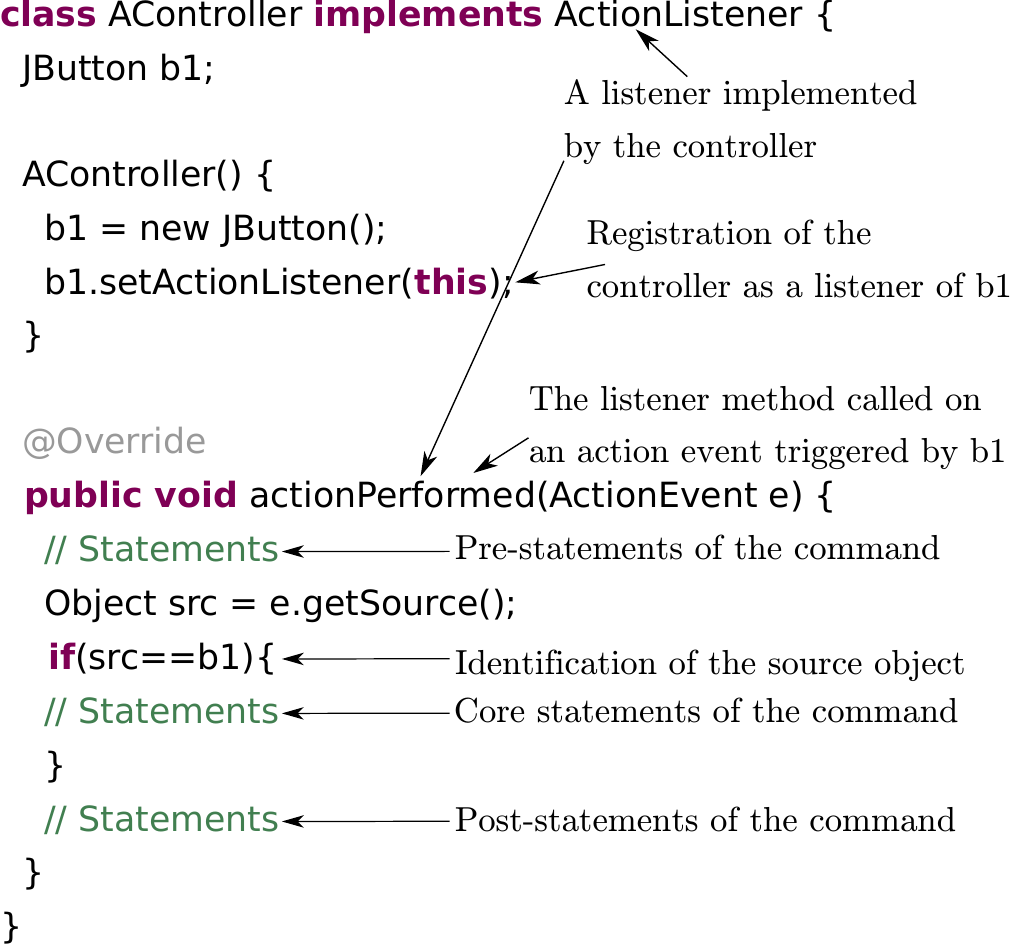}
	\caption{Composition of a UI listener and its UI commands}\label{fig.cmd}
\end{figure}

Programming languages and their different UI toolkits propose various ways to define UI listeners and commands.
\Cref{fig.cmd} depicts how UI listeners and commands can be defined in Java UI toolkits.
\Cref{fig.cmd} shows the Java Swing code of a controller, \emph{AController}, that manages one button called \emph{b1}.
To receive and treat the events produced by users when triggering \emph{b1}, the controller has to register a specific Java Swing listener called \emph{ActionListener} on \emph{b1}.
In this example, this listener is the controller itself that implements the \emph{ActionListener} interface.
The \emph{actionPerfomed} listener method, declared in \emph{ActionListener}, is called each time the user triggers \emph{b1} to produce a UI command.
Before executing the command, verifications may be done to check whether the command can be executed.
In our case, the source of the event is compared to \emph{b1} to check that \emph{b1} is at the origin of the event.
We call such conditional statements, \emph{source object identification statements}.
Then, statements, specified into such conditional statements and that compose the main body of the command, are executed.
Statements that may be defined before and after the main body of a command are also considered as part of the command.

UI listeners can be implemented in different ways, which complexifies their code analysis.
The three main ones are detailed as follows.

\textbf{Listeners as anonymous classes --} In \Cref{lst.good} listeners are defined as an anonymous class (\Crefrange{code.1e}{code.2e}) and register with one interactive object (\Cref{code.0e}).
The methods of this listener are then implemented to define the command to perform when an event occurs.
Because such listeners have to handle only one interactive object, \emph{if} statements used to identify the involved interactive object are not more used, simplifying the code.

\textbf{Listeners as lambdas --} \Cref{lst.goodJava8} illustrates the same code than \Cref{lst.good} but using Lambdas supported since Java 8.
Lambdas simplify the implementation of anonymous class that have a single method to implement.

\textbf{Listeners as classes --} In some cases, listeners have to manage different intertwined methods.
This case notably appears when developers want to combine several listeners or methods of a single listener to develop a more complex user interaction.
For example, \Cref{lst.listClass} is a code excerpt that describes a mouse listener where different methods are managed:
\emph{mouseClicked} (\Cref{code.0g}), \emph{mouseReleased} (\Cref{code.1g}), and \emph{mouseEntered} (\Cref{code.2g}).
Data are shared among these methods (\emph{isDrag}, \Cref{code.3g,code.4g}).

\begin{lstlisting}[language=MyJava, caption={UI listeners as anonymous classes}, label=lst.good]
private void registerInteractiveObjectHandlers() {
 view.resetPageButton().addActionListener(|\label{code.0e}|
   new ActionListener() {|\label{code.1e}|
   %%@Override%% public void actionPerformed(ActionEvent e) {
     requestData(pageSize, null);
   }
 });|\label{code.2e}|
 view.previousPageButton().addActionListener(
   new ActionListener() {
     %%@Override%% public void actionPerformed(ActionEvent e) {
       if(hasPreviousBookmark())
         requestData(pageSize, getPreviousBookmark());
     }
   });//...
}
\end{lstlisting}

\begin{lstlisting}[language=MyJava, caption={A UI listener defined as a class}, label=lst.listClass]
class IconPaneMouseListener implements MouseListener {
  %%@Override%% public void mouseClicked(MouseEvent e) {|\label{code.0g}|
    if(!isDrag) {|\label{code.3g}|
      //...
    }
  }
  %%@Override%% public void mouseReleased(MouseEvent e) {|\label{code.1g}|
    isDrag = false;|\label{code.4g}|
  }
  %%@Override%% public void mouseEntered(MouseEvent e) {|\label{code.2g}|
    isMouseExited = false; 
    // ...
}}
\end{lstlisting}

\begin{lstlisting}[language=MyJava, caption={Same code than in Listing~\ref{lst.good} but using Java 8 Lambdas}, label=lst.goodJava8]
private void registerInteractiveObjectHandlers() {
 view.resetPageButton().addActionListener(
              e -> requestData(pageSize, null));|\label{code.0f}|
   
 view.previousPageButton().addActionListener(e -> {
  if (hasPreviousBookmark())
    requestData(pageSize, getPreviousBookmark());
 });
 
 //...
}
\end{lstlisting}

\subsection{Detecting UI commands}

This section details the algorithm for statically detecting UI commands in source code.
\Cref{algo.detect} summarizes the UI command detection process. 
The detection process includes three main steps.
First, UI listeners are identified in the code.
This trivial step consists of collecting the classes that implements UI listener interfaces provided by the supported UI toolkits (\Cref{code.detect1}).
Second, from the body statements of each detected listener class, \emph{source object identification statements} are identified (\cf \Cref{sub.srcWidget}).
Third, the statements that compose the different commands managed in one listener are identified (\cf \Cref{sub.cmdStatmts}).

\begin{algorithm} 
\caption{UI command detection}\label{algo.detect}
\begin{algorithmic}[1]\small
\Require $classes$, the source classes of the software system
\Require $tkListeners$, the listener classes/interfaces of supported UI toolkits
\Ensure $cmds$, the detected UI commands
\State $cmds \gets \varnothing$
\State $uiListeners \gets findUIListeners(classes, tkListeners)$\label{code.detect1}
\ForAll{$l \in uiListeners$}
   \State $s \gets l.getBody()$
   \State $objectIdentifs \gets findSrcObjectIdentificationStatmts(s)$
   \If{$objectIdentifs == \varnothing$}                        \label{code.detect2}
      \State $cmds \gets cmds \cup \{Command(s, \varnothing\}$\label{code.detect3}
   \Else
      \ForAll{$i \in objectIdentifs$}
         \State $cmds \gets$
         \State ~~~~$cmds \cup \{extractCmdFromObjectIdentif(i, s)\}$
      \EndFor
   \EndIf	
\EndFor
\State
\Function{findSrcObjectIdentificationStatmts}{$statmts$}\label{code.detect4}
   \State $identifs \gets findConditionalStatmts(statmts)$
   \State $identifs \gets filterConditionalsThatUseUIEvent(identifs)$
   \State $identifs \gets filterLastNestedConditionals(identifs)$
   \State \Return $identifs$\label{code.detect5}
\EndFunction
\State
\Function{extractCmdFromObjectIdentif}{$conds$, $statmts$}\label{code.detect6}
   \State $s \gets conds.getStatmts()$
   \State $s \gets s \cup backwardSlicing(conds.getStatmts())$
   \State $s \gets s \cup forwardSlicing(conds.getStatmts())$
   \State $considerDispatchedMethods(s)$
   \State \Return $Command(s, conds)$\label{code.detect7}
\EndFunction
\end{algorithmic}
\end{algorithm}

\subsubsection{Identifying source object identification statements}\label{sub.srcWidget}

Software developers are free to develop UI listeners that can produce a single or multiple UI commands.
To identify the interactive object at the origin of a given event, conditional statements are commonly used.
We call such conditional statements, \emph{source object identification statements}.
Such conditional blocks may encapsulate a command to execute in reaction to the event.
Their identification is therefore mandatory to detect UI commands.
For example, five nested conditional blocks (\Cref{code.1v,code.3v,code.4v,code.6v,code.8v}) compose the listener method \emph{actionPerformed} in \Cref{lst.fitsAllSwing}.
The first conditional block checks the type of the interactive object that produced the event (\Cref{code.1v}):
\begin{lstlisting}[language=MyJava,numbers=none,escapechar=~]
if(src instanceof JMenuItem || src instanceof JButton)
\end{lstlisting}
This block contains three other conditional blocks that identify the interactive object using its action command (\Cref{code.3v,code.6v,code.8v}), for example:
\begin{lstlisting}[language=MyJava,numbers=none,escapechar=~]
if(cmd.equals("Copy"))
\end{lstlisting}
Each of these three blocks encapsulates one command to execute in reaction of the event.

As summarized in \Cref{algo.detect} (\Crefrange{code.detect4}{code.detect5}), the detection of source object identification statements starts by identifying all the conditional statements contained in the given listener method.
Then, only the conditional statements that directly or indirectly use the UI event given as a parameter of the listener method are considered.
For example \verb|if(selectedText)| (\Cref{code.4v}) makes no use of the UI event \emph{e} and is thus not considered as a source object identification statement.
Finally, on nested conditional statements, the last nested one is considered as the main source object identification of a command.

We empirically identified three kinds of source object identification statements by looking an existing Java code, as explained as follows.

\textbf{Comparing a property of the interactive object --}
\Cref{lst.fitsAllSwing} is an example of the first variant of source object identification statements: 
the interactive object that produced the event (\cref{code.3v,code.6v,code.8v}) are identified with a string value associated to the interactive object and returned by \emph{getActionCommand} (\cref{code.2v}).
Each of the three \emph{if} blocks forms a UI \emph{command} to execute in response of the event produced by interacting with a specific interactive object (\cref{code.4v,code.5v,code.7v,code.9v}).

\begin{lstlisting}[language=MyJava, caption={interactive object identification using interactive object's properties in Swing}, label=lst.fitsAllSwing,escapechar=~]
public class MenuListener 
			        implements ActionListener, CaretListener {
 protected boolean selectedText;

 %%@Override%% public void actionPerformed(ActionEvent e) {
  Object src = e.getSource();
  if(src instanceof JMenuItem || src instanceof JButton){~\label{code.1v}~ 
		 String cmd = e.getActionCommand(); ~\label{code.2v}~ 
		 if(cmd.equals("Copy")){~\label{code.3v}~//Command #1 
			 if(selectedText)~\label{code.4v}~
				 output.copy();~\label{code.5v}~
		 }else if(cmd.equals("Cut")){~\label{code.6v}~//Command #2
			  output.cut();~\label{code.7v}~
		 }else if(cmd.equals("Paste")){~\label{code.8v}~//Command #3
			 output.paste();~\label{code.9v}~
		 }
		 // etc.
		}~\label{code.15v}~
  }
  %%@Override%% public void caretUpdate(CaretEvent e){
   	selectedText = e.getDot() != e.getMark();
   	updateStateOfMenus(selectedText);	
}}
\end{lstlisting}

In Java Swing, the properties used to identify interactive objects are mainly the \emph{name} or the \emph{action command} of these interactive objects.
The action command is a string value used to identify the source interactive object using a UI event.
\Cref{lst.init}, related to \Cref{lst.fitsAllSwing}, shows how an action command (\cref{code.1b,code.3b}) and a listener (\cref{code.2b,code.4b}) can be associated to an interactive object in Java Swing during the creation of the user interface.

\begin{lstlisting}[language=MyJava, caption={Initialization of Swing interactive objects to be controlled by the same listener}, label=lst.init]
menuItem = new JMenuItem();
menuItem.setActionCommand("Copy");|\label{code.1b}|
menuItem.addActionListener(listener);|\label{code.2b}|

button = new JButton();
button.setActionCommand("Cut");|\label{code.3b}|
button.addActionListener(listener);|\label{code.4b}|
//...
\end{lstlisting}

\textbf{Checking the type of the interactive object --}
The second variant of source object identification statements consists of checking the \emph{type} of the interactive object that produced the event.
\Cref{lst.instanceof} depicts such a practice where the type of the interactive object is tested using the operator \emph{instanceof} (\Cref{code.1c,code.2c,code.3c,code.4c}).
One may note that such \emph{if} statements may have nested \emph{if} statements to test properties of the interactive object as explained in the previous point.

\begin{lstlisting}[language=MyJava, caption={interactive object identification using the operator \emph{instanceof}},label=lst.instanceof]
public void actionPerformed(ActionEvent evt) {
   Object target = evt.getSource();
   if (target instanceof JButton) {|\label{code.1c}|
      //...
   } else if (target instanceof JTextField) {|\label{code.2c}|
      //...
   } else if (target instanceof JCheckBox) {|\label{code.3c}|
      //...
   } else if (target instanceof JComboBox) {|\label{code.4c}|
      //...
}}
\end{lstlisting}

\textbf{Comparing interactive object references --}
The last variant consists of comparing interactive object references to identify those at the origin of the event.
\Cref{lst.fitsAllGWT} illustrates this variant where \emph{getSource} returns the source interactive object of the event that is compared to interactive object references contained by the listener (\eg \cref{code.1d,code.2d,code.3d}).

\begin{lstlisting}[language=MyJava, caption={Comparing interactive object references}, label=lst.fitsAllGWT]
public void actionPerformed(ActionEvent event) {
   if(event.getSource() == view.moveDown) {|\label{code.1d}|
      //...
   } else if(event.getSource() == view.moveLeft) {|\label{code.2d}|
      //...
   } else if(event.getSource() == view.moveRight) {|\label{code.3d}|
      //...
   } else if(event.getSource() == view.moveUp) {
      //...
   } else if(event.getSource() == view.zoomIn) {
      //...
   } else if(event.getSource() == view.zoomOut) {
      //...
   }}
\end{lstlisting}

In these three variants, multiple \emph{if} statements are successively defined.
Such successions are required when one single UI listener gathers events produced by several interactive objects.
In this case, the listener needs to identify the interactive object that produced the event to process.
When no source object identification statement is detected in a UI listener, all the statements of the listener are considered as part of a unique command (\Cref{code.detect2,code.detect3} in \Cref{algo.detect}).

The three variants of source object identification statements appear in all the main Java UI toolkits, namely Swing, SWT, GWT, and JavaFX.
Examples for these toolkits are available on the companion webpage of this paper\footnoteref{foot.webpage}.

\subsubsection{Extracting UI command statements}\label{sub.cmdStatmts}

We consider that each source object identification statement surrounds a UI command.
From a source object identification statement, the code statements that compose the underlying UI command are identified as follows (\Crefrange{code.detect6}{code.detect7}, \Cref{algo.detect}).
First, the statements that compose the conditional statement spotted as a source object identification statement are collected and considered as the main statements of the command.
Second, these statements may depend on variables or attributes previously defined and used.
A static backward slicing~\cite{XU05} is done to gather all these code elements and integrate them into the command.
For example, the following code illustrates the statements sliced for the first of the two commands of the listener.
The statements \cref{code.slice1,code.slice2} are sliced since they are used by the source object identification statements of command~\#1: \emph{s.equals("Copy")}
\begin{lstlisting}[language=MyJava]
public void actionPerformed(ActionEvent evt) {
 Object src = evt.getSource(); // sliced|\label{code.slice1}|
  if(src instanceof JMenuItem){ // Object identification
   String s = evt.getActionCommand(); // sliced|\label{code.slice2}|
	if(s.equals("Copy")){ // Cmd #1 Object identification
	 if(selectedText) // Main command statement
	  output.copy(); // Main command statement
	}else if(s.equals("Cut")){ // Cmd #2 Object identification
	 output.cut(); // Not considered in cmd #1|\label{code.slice4}|
	}
	output.done(); // sliced |\label{code.slice3}|
  }
}
\end{lstlisting}
Similarly, a static forward slicing is done to gather all the code elements related to the command defined after the source object conditional statement.
For example, the statement located \Cref{code.slice3} is sliced since \emph{output} is used by Command~\#1.
The statements part of the main statements of another command are not sliced (\eg \Cref{code.slice4}).

Some listeners do not treat the UI event but dispatch this process to another method, called \emph{dispatch method}.
The following code excerpt depicts such a case.
Each method invocation that takes as argument the event or an attribute of the event is considered as a dispatch method (\eg \Cref{code.slice5,code.slice6}).
In this case, the content of the method is analyzed similarly to the content of a listener method.
\begin{lstlisting}[language=MyJava]
public void actionPerformed(ActionEvent evt) {
 treatEvent(evt);|\label{code.slice5}|
}

private void treatEvent(AWTEvent evt) {
 Object src = evt.getSource();|\label{code.slice6}|
 //...
}
\end{lstlisting}

\begin{table*}[h]\small
 \caption{The four selected software systems and their characteristics}\label{tab.appSize} 
  \centering \setlength{\tabcolsep}{2.2pt}
    \begin{tabular}{p{2.5cm}llSSSp{8.0cm}}
    \toprule
\textbf{Software system}& \textbf{Version} & \textbf{UI toolkit} & \textbf{kLoCs} & \textbf{\# commits} & \textbf{\# UI listeners} &\textbf{Source repository link}\\
 	\cmidrule(lr){1-1}\cmidrule(lr){2-2}\cmidrule(lr){3-3}\cmidrule(lr){4-4} \cmidrule(lr){5-5} \cmidrule(lr){6-6} \cmidrule(lr){7-7} 
Eclipse \mbox{\scriptsize(platform.ui.workbench)} & 4.7 & SWT & 143 & 10049 & 259 & \scriptsize \url{https://git.eclipse.org/c/gerrit/platform/eclipse.platform.git/}\smallskip\\
JabRef & 3.8.0 & Swing & 95 & 8567 & 486 & \scriptsize\url{https://github.com/JabRef/jabref}\smallskip\\
ArgoUML & 0.35.1 & Swing & 101 & 10098 & 214 & \scriptsize\url{https://github.com/rastaman/argouml-maven}\\
 &  & & & & & \scriptsize\url{http://argouml.tigris.org/source/browse/argouml/}\smallskip\\
FreeCol & 0.11.6 & Swing & 118 & 12330 & 223 & \scriptsize\url{https://sourceforge.net/p/freecol/git/ci/master/tree/}\\
	\bottomrule
    \end{tabular}
\end{table*}

\section{An empirical study on UI listeners}\label{sec.study}

UI listeners are used to bind UIs to their underlying software system.
The goal of this study is to \emph{state whether the number of UI commands that UI listeners can produce has an effect on the code quality of these listeners}.
Indeed, software developers are free to develop UI listeners that can produce a single or multiple UI commands since no coding practices or UI toolkits enforce coding recommendations.
To do so, we study to what extent the number of UI commands that a UI listener can produce has an impact on the change- and fault-proneness of the UI listener code.
Such a correlation has been already studied to evaluate the impact of several antipatterns on the code quality~\cite{Khomh2012}.
Change- and fault-proneness are considered in the literature as negative impacts of a design smell on the code~\cite{palomba14,Khomh2012,lozano2007assessing,rapu2004using,Frolin15}.

We formulate the research questions of this study as follows:
\begin{itemize}\itemsep-0.1cm
  \item[\textbf{RQ1}] To what extent the number of UI commands per UI listeners has an impact on fault-proneness of the UI listener code?
  \item[\textbf{RQ2}] To what extent the number of UI commands per UI listeners has an impact on change-proneness of the UI listener code?
  \item[\textbf{RQ3}] Do developers agree that a threshold value, \ie a specific number of UI commands per UI listener, that can characterize a UI design smell exist?
\end{itemize}

To answer these three research questions, we measured the following independent and dependent variables.
All the material of the experiments is freely available on the companion web page\footnoteref{foot.webpage}.

\subsection{Tool}\label{sec.inspector}

The command detection algorithm has been implemented in \tool, an open-source Eclipse plug-in that analyzes Java Swing, SWT, and JavaFX software systems\footnoteref{foot.webpage}.
\tool uses \emph{Spoon}, a library for transforming and analyzing Java source code~\cite{spoon}, to support the static analyses.

\subsection{Independent Variables}

\noindent\textbf{Number of UI Commands} (\textbf{CMD}).
This variable measures the number of UI commands a UI listener can produce.
To measure this variable, we use the proposed static code analysis algorithm detailed in \Cref{sec.cmd} and implemented in \tool.

\subsection{Dependent Variables}\label{sub.depvars}

\noindent\textbf{Average Commits} (\textbf{COMMIT}).
This variable measures the average number of commits of UI listeners.
This variable will permit to evaluate the change-proneness of UI listeners.
To measure this variable, we automatically count the number of the commits that concern each UI listener.

\medskip

\noindent\textbf{Average fault Fixes} (\textbf{FIX}). 
This variable measures the average number of fault fixes of UI listeners.
This variable will permit to evaluate the fault-proneness of UI listeners.
To measure this variable, we manually analyze the log of the commits that concern each UI listener.
We manually count the commits which log refers to a fault fix, \ie logs that point to a bug report of an issue-tracking system (using a bug ID or a URL) or that contain the term "\emph{fix}" (or a synonymous).
We use the following list of terms to identify a first list of commits:
\emph{fix, bug, error, problem, work, issue, ticket, close, reopen, exception, crash, NPE, IAE, correct, patch, repair, rip, maintain, warning}.
We then manually scrutinized each of these commits.
For example, the following commit message extracted from the Eclipse history is considered as a bug fix as it fixes a Java exception (\emph{NullPointerException}, \emph{NPE}):
"\emph{49216 [About] NPE at ..AboutFeaturesDialog.buttonPressed}".
However, we do not consider the following commit "\emph{Bug 509477 - Use lambdas in ...ui.workbench"} as a bug fix.
The term "bug" is used here as a reference to the issue tracking system and not to a real error.

Both COMMIT and FIX rely on the ability to get the commits that concern a given UI listener.
For each software system, we use all the commits of their history as the time-frame of the analysis.
We ignore the first commit as it corresponds to the creation of the project.

The size, \ie the number of lines of code (LoC), of UI listeners may have an impact on the number of commits and fault fixes.
So, we need to compare UI listeners that have a similar size by computing the four quartiles of the size distribution of the UI listeners~\cite{Abbes2011,aniche2017}.
We kept the fourth quartile ($Q_4$) as the single quartile that contains enough listeners with different numbers of commands to conduct the study.
This quartile $Q_4$ contains 297 UI listeners that have more than 10 lines of code.
For the study the code has been formatted and the blank lines and comments have been removed.

Commits may change the position of UI listeners in the code (by adding or removing LoCs).
To get the exact position of a UI listener while studying its change history, we use the Git tool \emph{git-log}\footnote{\scriptsize\url{https://git-scm.com/docs/git-log}}.
The \emph{git-log} tool has options that permit to: 
follow the file to log across file renames (option~\mbox{\emph{-M}});
trace the evolution of a given line range across commits (option~\mbox{\emph{-L}}).
We then manually check the logs for errors.

\begin{table*}[t]\small
  \centering\setlength{\tabcolsep}{4pt}
  \caption{Means, correlations, and Cohen's \emph{d} of the results}\label{tab.results}
    \begin{tabular}{cSSScccc}
    \toprule
\textbf{Dependent} & \textbf{Mean} & \textbf{Mean} & \textbf{Mean}  & \textbf{Correlation} & \textbf{Cohen's d} & \textbf{Cohen's d} & \textbf{Cohen's d} \\
\textbf{variables} & \textbf{CMD=1} & \textbf{CMD=2} & \textbf{CMD>=3} & \textbf{(significance)} & \textbf{CMD=1 vs CMD=2} & \textbf{CMD=2 vs CMD>=3} & \textbf{CMD=1 vs CMD>=3}\\
                  &                  &                &               &                       & \textbf{(significance)} & \textbf{(significance)} & \textbf{(significance)}\\
 \cmidrule(lr){1-1} \cmidrule(lr){2-2} \cmidrule(lr){3-3} \cmidrule(lr){4-4} \cmidrule(lr){5-5} \cmidrule(lr){6-6} \cmidrule(lr){7-7} \cmidrule(lr){8-8}
$FIX$    & 1.107   & 1.149  & 2.864  & 0.4281   &  0.0301  & 0.5751  & 0.8148 \\
         &         &        &        & (***) & (no)   & (no)  & (***)  \\\midrule
$COMMIT$ & 5.854   & 6.872  & 10.273 & 0.3491   & 0.1746  & 0.3096 & 0.5323 \\
         &         &        &        & (***) & (no) & (no)  & (no)\\\bottomrule
    \end{tabular}
\end{table*}

\subsection{Objects}

The objects of this study are open-source software systems.
The dependent variables, previously introduced, impose several constraints on the selection of these software systems.
They must use an issue-tracking system and the Git version control system.
We focused on software systems that have more than \num{5000} commits in their change history to let the analysis of the commits relevant.
In this work, we focused on Java Swing and SWT UIs because of the popularity and the large quantity of Java Swing and SWT legacy code available on code repositories such as \emph{Github}\footnote{\scriptsize\url{https://github.com/}} and \emph{Sourceforge}\footnote{\scriptsize\url{https://sourceforge.net/}}.
We thus selected four Java Swing and SWT software systems, namely ArgoUML, JabRef, Eclipse (more precisely the \emph{platform.ui.workbench} plug-in), and Freecol.
\Cref{tab.appSize} lists these software systems, the version used, their UI toolkit, their number of Java line of codes, commits, and UI listeners, and the link the their source code.
The number of UI listeners excludes empty listeners.
The average number of commits of these software systems is approximately \num{10.2}k commits.
The total size of Java code is \num{457}k Java LoCs, excluding comments and blank lines.
Their average size is approximately \num{114}k Java LoCs.

\subsection{Results}

We can first highlight that the total number of UI listeners producing at least one UI command identified by our tool is \num{1205}, \ie an average of \num{301} UI listeners per software system.
This approximately corresponds to \num{11}~kLoCs of their Java code.

\begin{figure}[h]
	\centering
		\includegraphics[width=0.83\columnwidth]{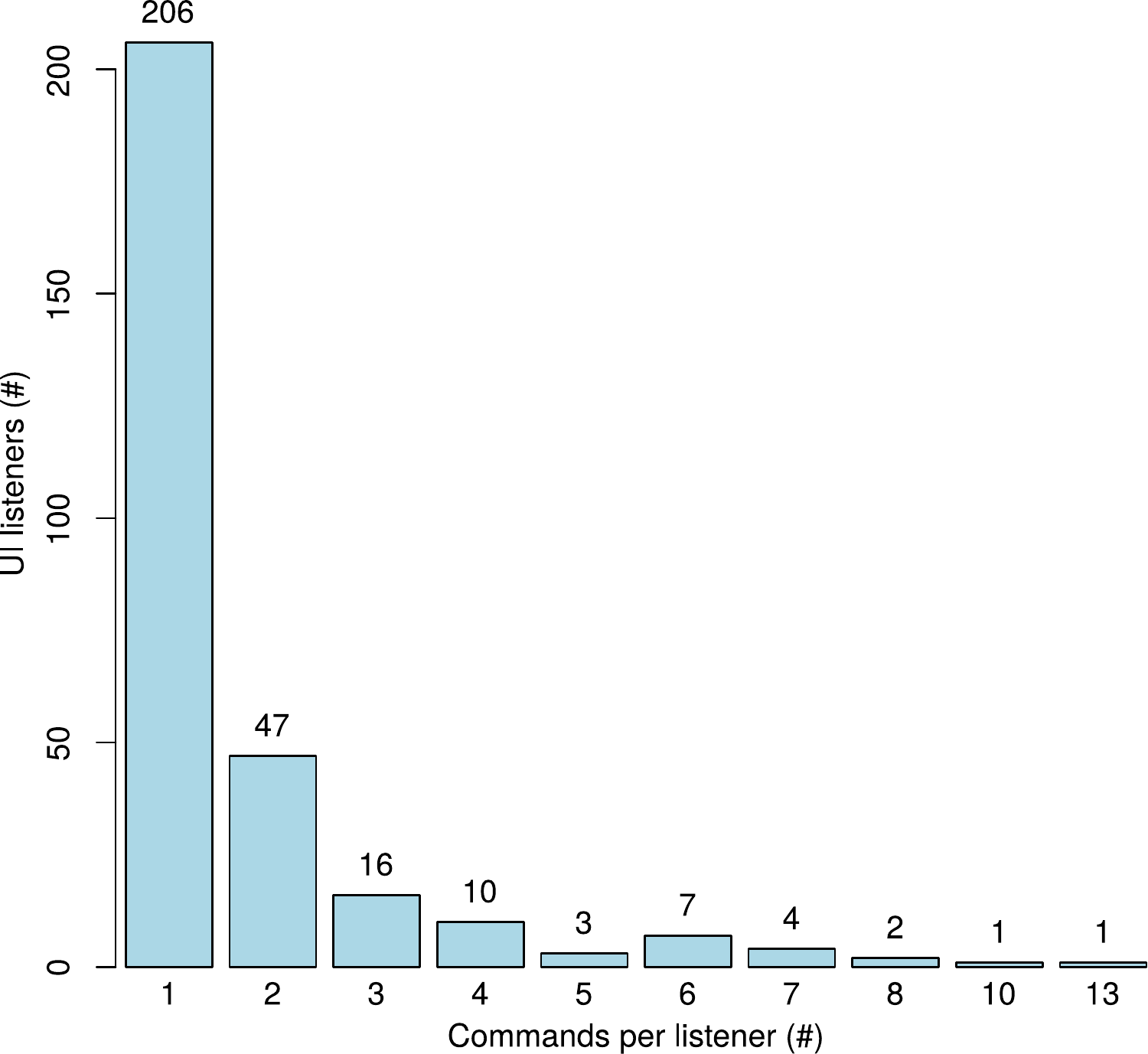}
		\caption{Distribution of the listeners according to their number of UI commands}\label{fig.nbCmds}
\end{figure}

As explained in \Cref{sub.depvars}, to compare listeners with similar sizes we used the quartile $Q_4$ for the study.
\Cref{fig.nbCmds} shows the distribution of the listeners of $Q_4$ according to their number of UI commands.
\SI{69.36}{\percent} (\ie \num{206}) of the listeners can produce one command (we will call them one-command listeners).
\SI{30.64}{\percent} of the listeners can be produce two or more commands:
\num{47} listeners can produce two commands.
\num{16} listeners can produce three commands.
\num{28} listeners can produce at least four commands.
To obtain representative data results, we considered in the following analyses three categories of listeners: 
\emph{one-command listener} (\emph{CMD=1} in \Cref{tab.results}), \emph{two-command listener} (\emph{CMD=2}), \emph{three+-command listener} (\emph{CMD>=3}).

We computed the means of \emph{FIX} and \emph{COMMIT} for each of these three categories.
To compare the effect size of the means (\ie \emph{CMD=1 vs. CMD=2}, \emph{CMD=1 vs CMD=2}, \emph{and CMD=1 vs. CMD>=3}) we used the Cohen's \emph{d} index~\cite{Sheskin2007}.
Because we compared multiple means, we used the Bonferroni-Dunn test~\cite{Sheskin2007} to adapt the confidence level we initially defined at \SI{95}{\percent} (\ie $\alpha=0.05$):
we divided this $\alpha$ level by the number of comparisons (3), leading to $\alpha=0.017$.
We used the following code scheme to report the significance of the computed $p$-value:
No significance= $p>0.017$, $*= p\le .0017$, **= $p\le .005$, $***=p\le .001$.
Because \emph{FIX} (resp. \emph{COMMIT}) and \emph{CMD} follow a linear relationship, we used the Pearson's correlation coefficient to assess the correlation between the number of fault fixes (resp. number of changes) and the number of UI commands in UI listeners~\cite{Sheskin2007}.
The correlation is computed on all the data of \Cref{fig.nbCmds} (\ie not using the three categories of listeners).
The results of the analysis are detailed in \Cref{tab.results}.

\begin{figure}[h]
	\centering
		\includegraphics[width=0.97\columnwidth]{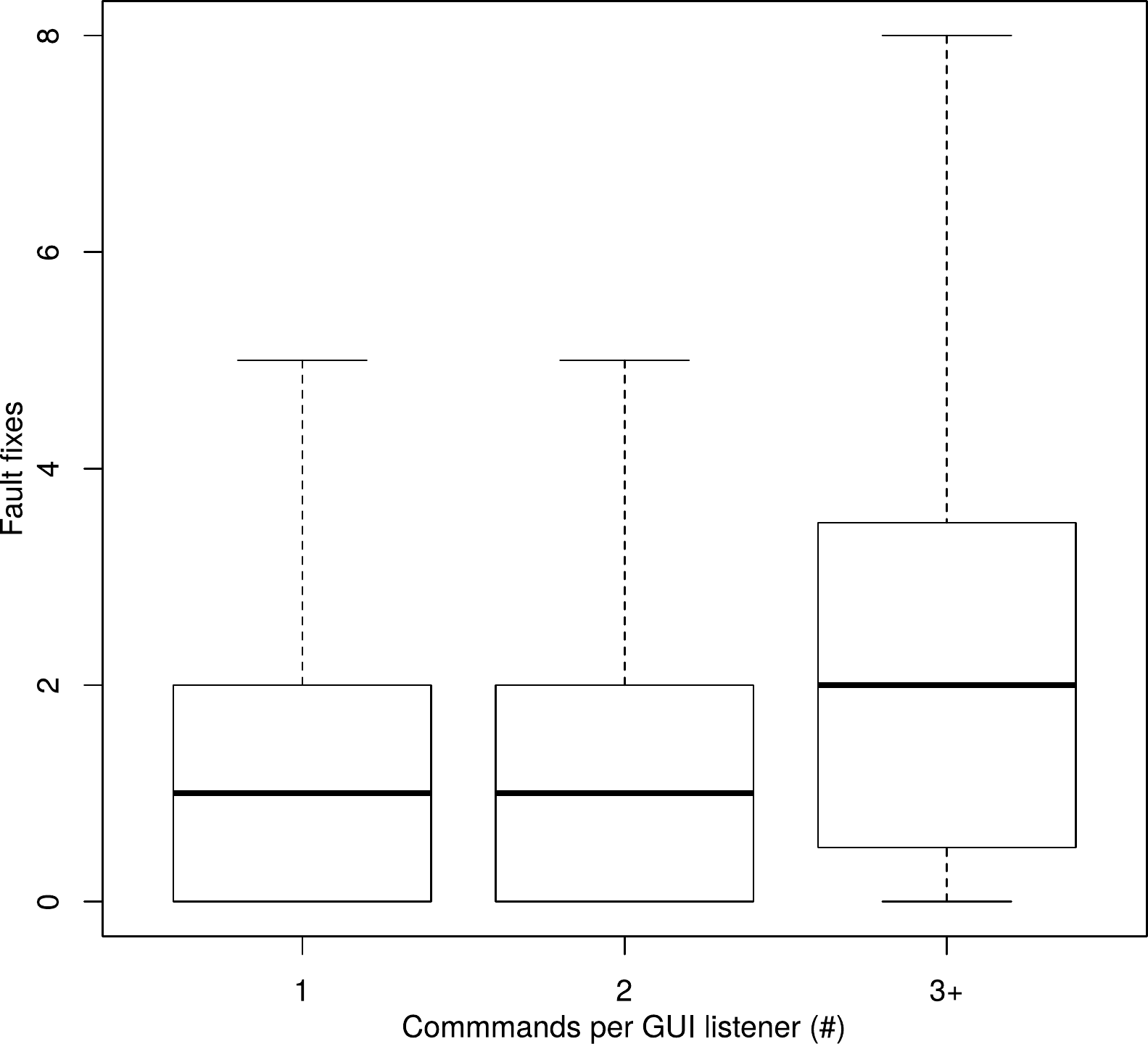}
		\caption{Number of fault fixes of UI listeners}\label{fig.bugs}
\end{figure}

\Cref{fig.bugs} depicts the evolution of \emph{FIX} over \emph{CMD}.
We observe a significant increase of the fault fixes when $CMD\ge 3$.
According to the Cohen's \emph{d} test, this increase is large (\num{0.8148}).
\emph{FIX} increases over \emph{CMD} with a moderate correlation (\num{0.4281}, if in $[0.3, 0.7[$, a correlation is considered to be moderate~\cite{Sheskin2007}).

Regarding \textbf{RQ1}, on the basis of these results we can conclude that managing several UI commands per UI listener has a negative impact on the fault-proneness of the UI listener code:
a significant increase appears at three commands per listener, compared to one-command listeners.
There is a moderate correlation between the number of commands per UI listener and the fault-proneness.

\begin{figure}[h]
	\centering
		\includegraphics[width=0.99\columnwidth]{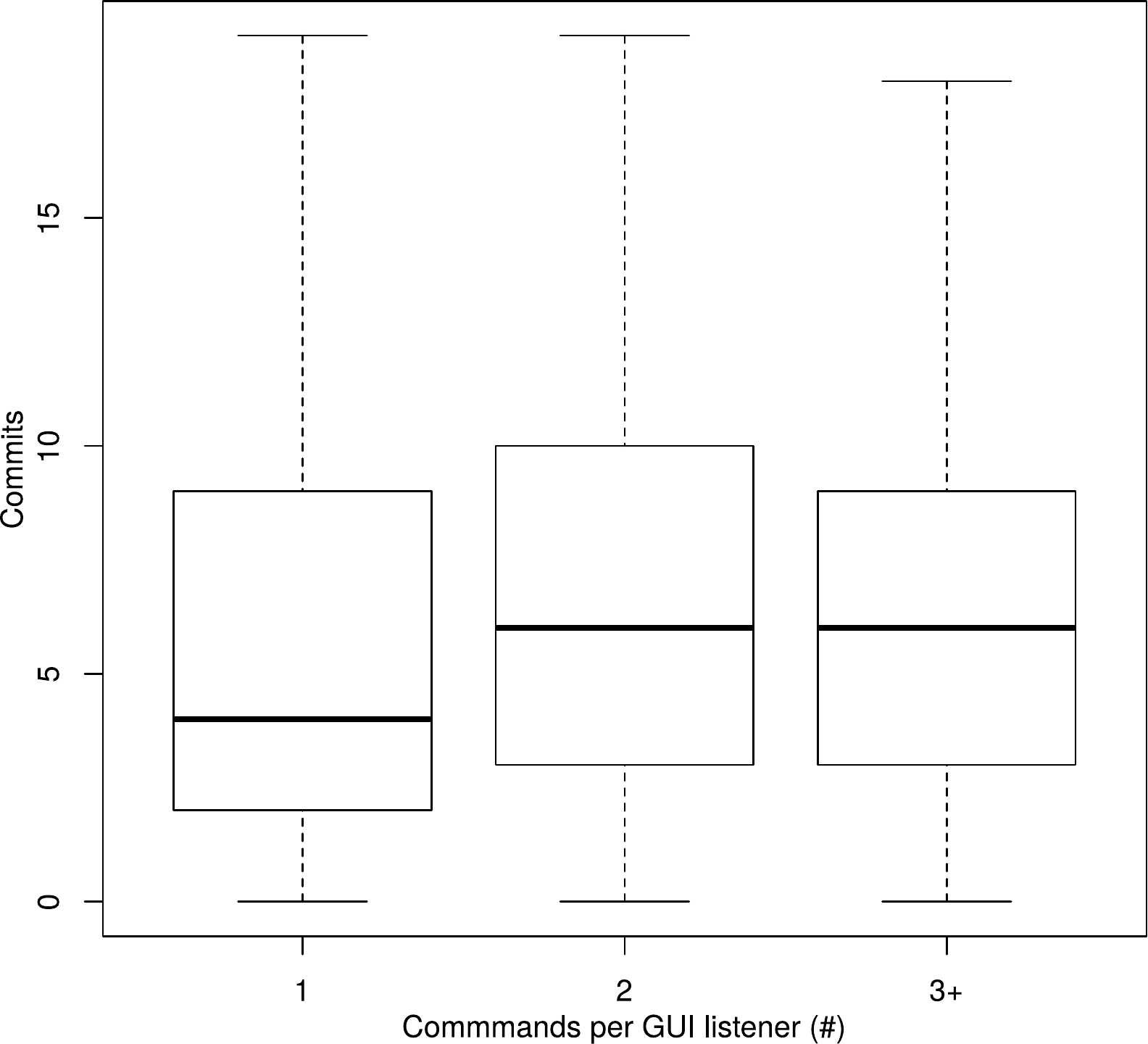}
		\caption{Number of commits of UI listeners}\label{fig.commits}
\end{figure}

\bigskip
\Cref{fig.commits} depicts the evolution of \emph{COMMIT} over \emph{CMD}.
The mean value of \emph{COMMIT} increases over \emph{CMD} with a weak correlation (\num{0.3491}, using the Pearson's correlation coefficient).
A medium (Cohen's \emph{d} of 0.5323) but not significant ($p$-value of \num{0.0564}) increase of COMMIT can be observed between one-command and three+-command listeners.
\emph{COMMIT} increases over \emph{CMD} with a moderate correlation (\num{0.3491}).

Regarding \textbf{RQ2}, on the basis of these results we can conclude that managing several UI commands per UI listener has a small but not significant negative impact on the change-proneness of the UI listener code.
There is a moderate correlation between the number of commands per UI listener and the change-proneness.

\bigskip
Regarding \textbf{RQ3}, we observe a significant increase of the fault fixes for three+-command listeners against one-command listeners.
We also observe an increase of the commits for three+-command listeners against one-command listeners.
We thus state that a threshold value, \ie a specific number of UI commands per UI listener, that characterizes a UI design smell exists.
Note that since the COMMIT metrics counts all the commits, bug-fix commits included, the increase of the commits may be correlated to the increase of the fault fixes for three+-command listeners.
We contacted developers of the analyzed software systems to get feedback about a threshold value.
Beyond the "\emph{sounds good}" for three commands per listener, one developer explained that "\emph{strictly speaking I would say, more than one or two are definitely an indicator.
However, setting the threshold to low \emph{[lower than three commands per listener]} could lead to many false positives}".
Another developer said "\emph{more than one [command per listener] could be used as threshold, but generalizing this is probably not possible}".
We agree and \emph{define the threshold to three UI commands per UI listener}.
Of course, this threshold value is an indication and as any design smell it may vary depending on the context.
Indeed, as noticed in several studies, threshold values of design smells must be customizable to let system experts the possibility to adjust them~\cite{Johnson2013,palomba2014}.

The threats to validity of this empirical study are discussed in \Cref{sub.threats}.

\subsection{Introducing the Blob Listener design smell}\label{sec.blob}

Based on the results of the empirical study previously detailed, we showed that a significant increase of the fault fixes and changes for two- and three+-command listeners is observed.
Considering the feedback from developers of the analyzed software systems, we define at three commands per listener the threshold value from which a design smell, we called \bl, appears.
We define the \bl as follows:

\begin{definition}[Blob Listener]\label{def.blob}
A \bl is a UI listener that can produce several UI commands.
\bls can produce several commands because of the multiple interactive objects they have to manage.
In such a case, \bls' methods (such as \emph{actionPerformed}) may be composed of a succession of conditional statements that:
1) identify the interactive object that produced the UI event to treat;
2) execute the corresponding UI command.
\end{definition}

\begin{figure*}[th]
	\centering
		\includegraphics[width=1.95\columnwidth]{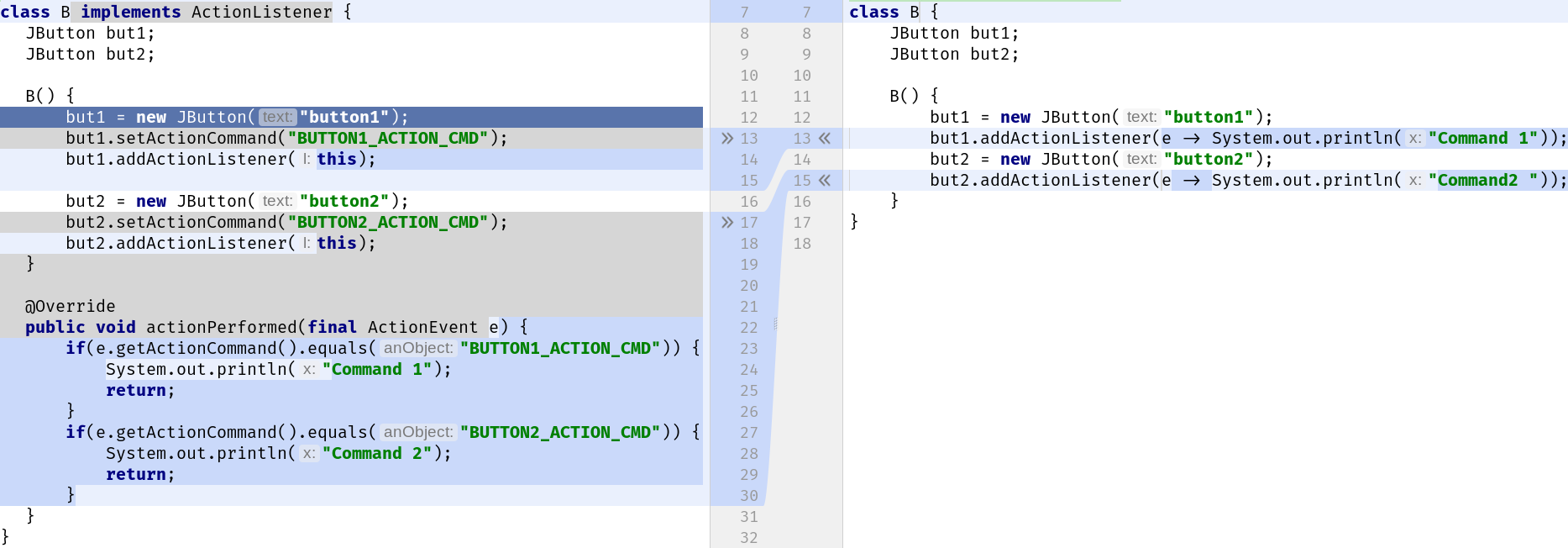}
		\caption{Simple example of the refactoring effects. On the left, the original source code. On the right the refactored source code}\label{fig.refactor}
\end{figure*}

\section{Refactoring \bls}\label{sec.refactor}

This section details the semi-automatic and behavior preserving code refactoring technique~\cite{Mens2004,fowler1999} for removing \bls from source code.
\Cref{algo.refactor} details the refactoring process.
The general idea of the refactoring is to move each command that compose a \bl into a new UI listener directly applied on the targeted interactive object.
\Cref{fig.refactor} illustrates the refactoring using a mere example.
The original UI listener of this example (Lines 22-30, on the left) manages two commands that can be produced by one interactive object each, namely \emph{but1} and \emph{but2}.
These two interactive objects register the listener Lines~14 and~18.
String values are used to identify the interactive object at the origin of the UI event (Lines~13, 17, 23, and~27).
Each command that composes a \bl (identified using the static code analysis detailed in \Cref{sec.cmd}) are moved into a new UI listener directly at the listener registration of the interactive object (Lines~13 and~15, on the right), as detailed in the following algorithm.

\begin{algorithm} 
\caption{\bl refactoring}\label{algo.refactor}
\begin{algorithmic}[1]\small
\Require $classes$, the source classes of the software system
\Require $blobs$, The \bl spotted in the source code
\State $allObjs \gets findAllInteractiveObjects(classes)$ 
\ForAll{$blob \in blobs$}
   \ForAll{$cmd \in blob.getCommands()$}
      \State $interObjects \gets findAssociatedInterObjects(cmd, allObjs)$
      \ForAll{$obj \in interObjects$}
         \State $registration \gets findListenerRegistration(obj)$
         \State $removeUselessStatements(cmd)$
         \State $newListener \gets new~UI~Listener$\label{algoRef.1}
         \State $replaceListenerRegistration(registration, newListener)$\label{algoRef.3}
         \State $copyStatements(cmd.getStatmts(), newListener)$\label{algoRef.2}
         \State $removeOldCommand(cmd, obj)$\label{algoRef.4}
      \EndFor
   \EndFor
\EndFor
\end{algorithmic}
\end{algorithm}

The first step of the algorithm identifies all the interactive objects declared in the source code and their usages (\Cref{sub.allWidgets}).
Then, the identification of the interactive objects associated to (\ie that can produce) each command of a \bl is done (\Cref{sub.assoWidgets}).
Following, for each associated interactive object found, a new UI listener object is created using the command statements.
This new listener is then directly used at the listener registration of the interactive object (\Cref{sub.createListener}).
Finally, the \bl is removed (\Cref{sub.removeblob}).

\subsection{Finding all the interactive objects and their usages}\label{sub.allWidgets}

As illustrated with \Cref{lst.fitsAllSwing,lst.init}, the definition of a \bl (\eg \Cref{lst.fitsAllSwing}) and its registrations to different interactive objects (\eg \Cref{lst.init}) are separated in the code.
Refactoring \bls first requires the identification of the interactive objects that register to the \bls.
To do so, a static code analysis scrutinizes the code to identify local variables and class attributes that are instances of UI toolkits interactive objects.
Then, the initialization statements of each interactive object, that we call \emph{interactive object usages}, are identified.
\begin{lstlisting}[language=MyJava,numbers=none]
menuItem = new JMenuItem();
menuItem.setActionCommand("COPY");
menuItem.addActionListener(listener);
\end{lstlisting}
For example the three statements of the code above are considered as interactive objects usages of \emph{menuItem} since they initialize these statements.
In the method that initializes a given interactive object, we consider all the statements using this interactive object as initialization statements.
The last initialization statement of these three lines of code is the registration of a UI listener (\emph{addActionListener}) to \emph{menuItem}.
Such a statement permits the identification of the interactive objects involved in the different UI listeners.
The next sub-section details how for each command of a given UI listener, interactive objects can be precisely identified.

\subsection{Finding associated interactive objects}\label{sub.assoWidgets}

\Cref{sec.cmd} details how the statements and the source object identification statements that compose a UI command are identified.
Based on the source object identification statements of each interactive object, we developed a static code analysis for identifying the potential interactive objects that may trigger the command among all the interactive objects of the software system.
As detailed in \Cref{sub.srcWidget}, we identified three kinds of source object identification statements.
The proposed analysis considers these three kinds to precisely identify the underlying interactive objects plus another technique, as explained as follows.

\medskip
\noindent\textbf{Comparing interactive object references --} 
Interactive object variables are compared to the object at the origin of the UI event, for example:
\begin{lstlisting}[language=MyJava,numbers=none,escapechar=~]
if(event.getSource() == view.moveDown)
\end{lstlisting}
In this simple case, we collect all the interactive objects used in the source object identification statements.

\medskip
\noindent\textbf{Comparing a property of the interactive object --}
The identification of the source interactive object may use some of its properties, such as its name or its action command.
For example a string literal can be used to identify an interactive object:
\begin{lstlisting}[language=MyJava,numbers=none,escapechar=~]
if(event.getActionCommand().equals("COPY"))
\end{lstlisting}
The analysis searches for string literals used both in the source object identification statements of a command and the usages of all the interactive objects.
The following code excerpt illustrates this step where \emph{button} uses the string value \emph{"COPY"}:
\begin{lstlisting}[language=MyJava,numbers=none,escapechar=~]
button.setActionCommand("COPY");
\end{lstlisting}
The use of string constants is also supported, for example:
\begin{lstlisting}[language=MyJava,numbers=none,escapechar=~]
public final String ACTION_CMD_COPY = "COPY";
//...
if(event.getActionCommand().equals(ACTION_CMD_COPY)) //...
\end{lstlisting}

\medskip
\noindent\textbf{Checking the type of the interactive object --} 
Checking the type of the source object that produced the event can help in detecting the underlying interactive objects.
The following source object identification statement implies that the underlying interactive object is a \emph{JButton}.
\begin{lstlisting}[language=MyJava,numbers=none,escapechar=~]
if(event.getSource() instanceof JButton)
\end{lstlisting}
This technique helps in reducing the number of candidates but may not precisely identify the underlying interactive object if several interactive objects of the same type are managed by the same \bl.

\medskip
\noindent\textbf{Analyzing listener registrations --}
As explained in \Cref{sec.cmd}, UI listener can be directly implemented during their registration to an interactive object.
The following code excerpt illustrates this case where the listener is implemented as an anonymous class.
\begin{lstlisting}[language=MyJava,numbers=none]
view.resetPageButton().addActionListener(
   new ActionListener() {
   %%@Override%% public void actionPerformed(ActionEvent e) {
     requestData(pageSize, null);
   }
});
\end{lstlisting}
In this case, the interactive object can be directly identified by looking at the registration method invocation.
However, since this practice permits the registration of a listener to a unique interactive object, it does not fit our case since \bls manage several interactive objects.

\bigskip

Once the involved interactive objects identified for each command, they are checked against the interactive objects that register the UI listener (\Cref{sub.allWidgets}).
This step aims at checking that the interactive objects used by a command and the interactive objects that register the UI listener of the command are the same.
Then, the refactoring is applied as detailed in the next section.
If no interactive object is identified neither in the listener registrations nor in the command statements, a warning is raised and the refactoring stopped.

\subsection{Fragmenting a \bl in atomic UI listeners}\label{sub.createListener}

As illustrated by \Cref{fig.refactor}, the goal of the refactoring process is to fragment a \bl into several atomic UI listeners directly defined at the listener registration (one atomic UI listener for each associated interactive object).
To do so, the developer can choose to either refactor \bls as Lambdas (concise and lightweight but requires Java~8) or as anonymous classes (more verbose but supported by all the Java versions).
As summarized in \Cref{algo.refactor}, a new UI listener is created for each interactive object of each UI command of a given \bl (\Cref{algoRef.1}, \Cref{algo.refactor}).
Each new atomic UI listener registers their targeted interactive object (\Cref{algoRef.3}, \Cref{algo.refactor}) as follows:
\begin{lstlisting}[language=MyJava,numbers=none]
interactiveObject.addActionListener(evt -> {
   // Command statements
});
\end{lstlisting}

Then, the statements that form the UI command are copied into this new atomic UI listener (\Cref{algoRef.2}, \Cref{algo.refactor}):
\begin{lstlisting}[language=MyJava,numbers=none]
interactiveObject.addActionListener(evt -> {
   output.copy();
});
\end{lstlisting}
All the source object identification statements (\eg Lines~23 and~27 in \Cref{fig.refactor}) are removed from the command statements since they are no more necessary.
Statements, such as \emph{return} statements (\eg Lines~25 and~29 in \Cref{fig.refactor}) that may end a UI command are also removed.
In some cases, the refactoring cannot be automatic and requires decisions from the developer, as discussed in \Cref{sub.limits}.

In some cases, several interactive objects can be associated to a single UI command.
For example, the following code excerpt shows that two different interactive objects \emph{button} and \emph{menu} (\Crefrange{twoWidgets.1}{twoWidgets.2}) perform the same command (\Cref{twoWidgets.3}).
\begin{lstlisting}[language=MyJava]
class Controller implements ActionListener {
   static final String ACTION_CMD = "actionCmd";
   JButton button;
   JMenuItem menu;

   Controller() {
      //...
      button.setActionCommand(ACTION_CMD);|\label{twoWidgets.1}|
      menu.setActionCommand(ACTION_CMD);|\label{twoWidgets.2}|
   }

   public void actionPerformed(ActionEvent e) {
      if(ACTION_CMD.equals(e.getActionCommand())) {|\label{twoWidgets.3}|
         // A single command for 'button' and 'menu'
	   }
   }
}
\end{lstlisting}
In such a case, the refactoring process creates one constant attribute that contains the shared UI listener (\Cref{same.1} in the following code excerpt).
Each interactive object registers the same listener (\Cref{same.2,same.3}).
This techniques does not contradict the \bl design smell (\ie UI listeners that can produce several commands) since the same command is produced by several interactive objects.
\begin{lstlisting}[language=MyJava]
class Controller {
   JButton button;
   JMenuItem menu;
   
   final ActionListener listener = e -> {/* The command */};|\label{same.1}|

   Controller() {
      button.setActionListener(listener);|\label{same.2}|
      menu.setActionListener(listener);|\label{same.3}|
   }
}
\end{lstlisting}

\subsection{Removing a \bl}\label{sub.removeblob}

Once a \bl fragmented into several atomic UI listeners, the \bl is removed from the source code.
As illustrated by \Cref{fig.refactor}, the UI listener method that forms the \bl is removed (here, \emph{actionPerformed}).
The class that contained the \bl does not implement the UI listener interface anymore (here, \emph{ActionListener}).
Statements of the initialization of the involved interactive objects may be removed if no more used and if used to identify the source interactive object in the \bl.
For Java Swing, such statements are typically invocations of the \emph{setActionCommand} method.

\subsection{Limits of the refactoring process}\label{sub.limits}

In some specific cases the refactoring cannot be done automatically.
For example, the following code excerpt shows a UI command (\Cref{codeAttr.4}) that makes use of a class attribute (\Cref{codeAttr.1}) initialized \Crefrange{codeAttr.2}{codeAttr.3}.
This listener does not define and register the interactive objects.
This job is done in another class.
\begin{lstlisting}[language=MyJava]
class Controller implements ActionListener {
	static final String ACTION_CMD = "actionCmd";
	JFileChooser c;|\label{codeAttr.1}|

	public Controller() {
		c = new JFileChooser();|\label{codeAttr.2}|
		c.setMultiSelectionEnabled(false);|\label{codeAttr.3}|
	}
	
   public void actionPerformed(ActionEvent e) {
      if(ACTION_CMD.equals(e.getActionCommand())) {
         c.showDialog(null, "title");|\label{codeAttr.4}|
	   }
   }
}
\end{lstlisting}
By default the class attributes (here \emph{c} and its initialization statements) used in UI commands are copied in the class that contains the interactive objects, as shown in the following refactored code that made use of the previous listener:
\begin{lstlisting}[language=MyJava]
class A {
	JFileChooser c;
   JButton b;

	public A() {
		c = new JFileChooser();
		c.setMultiSelectionEnabled(false);
		b = new JButton();
		b.setActionListener(e -> c.showDialog(null, "title"));
	}
}
\end{lstlisting}
This strategy may not be relevant in certain situations and the developer has to validate this change or finalize the refactoring.
We did not face this situation during our experiments of \Cref{sec.eval}.

\section{Evaluation}\label{sec.eval}

To evaluate the efficiency of our detection algorithm and refactoring process, we addressed the four following research questions:
\begin{itemize}\itemsep0cm
  \item[\textbf{RQ4}] To what extent is the detection algorithm able to detect UI commands in UI listeners correctly?
  \item[\textbf{RQ5}] To what extent is the detection algorithm able to detect \bls correctly?
  \item[\textbf{RQ6}] To what extent does the refactoring process propose correct refactoring solutions?
  \item[\textbf{RQ7}] To what extent the concept of \bl and the refactoring solution we propose are relevant?
\end{itemize}

The evaluation has been conducted using \tool, our implementation of the \bl detection and refactoring algorithms introduced in \Cref{sec.inspector}.
\tool leverages the Eclipse development environment to raise warnings in the Eclipse Java editor on detected Blob listeners and their UI commands.
The refactoring process is performed outside the Eclipse environment.
Initial tests have been conducted on software systems not reused in this evaluation.
\tool allows the setting of this threshold value to let system experts the possibility to adjust them, as suggested by several studies~\cite{Johnson2013,palomba2014}.
For the evaluation, the threshold has been set to two-commands per listener.
\tool and all the material of the evaluation are freely available on the companion web page\footnoteref{foot.webpage}.

\subsection{Objects}

We conducted our evaluation using the four large open-source software systems detailed in \Cref{sec.study}.

\subsection{Methodology}
The accuracy of the static analyses that compose the detection algorithm is measured by the \emph{recall} and \emph{precision} metrics~\cite{Frolin15}.
We ran \tool on each software system to detect UI listeners, commands, and \bls.
We assumed as a precondition that only UI listeners are correctly identified by our tool.
Thus, to measure the precision and recall of our automated approach, we manually analyzed all the UI listeners detected by \tool to:

\medskip
\noindent\textbf{Check commands}. We manually analyzed each UI listeners to state whether the UI commands they contain are correctly identified.
The \emph{recall} measures the percentage of correct UI commands that are detected (\Cref{eq.recallCmd}). 
The \emph{precision} measures the percentage of detected UI commands that are correct (\Cref{eq.precCmd}).
For 39 listeners, we were not able to identify the commands of UI listeners.
We removed these listeners from the data set.

\begin{equation}\textstyle\label{eq.recallCmd}\small
recall_{cmd} (\%) = \frac{|\{ correctCmds\}  \cap \{ detectedCmds\} |}{|\{ correctCmds\} |} \times 100   
\end{equation}

\begin{equation}\textstyle\label{eq.precCmd}\small
precision_{cmd} (\%) = \frac{|\{ correctCmds\} \cap \{ detectedCmds\}|}{|\{ detectedCmds\}|} \times 100
\end{equation}

The \emph{correctCmds} variable corresponds to all the commands defined in UI listeners, \ie the commands that should be detected by \tool.
The \emph{recall} and \emph{precision} are calculated over the number of false positives (FP) and false negatives (FN).
A UI command incorrectly detected by \tool is classified as false positive.
A false negative is a UI command not detected by \tool.

\medskip\noindent\textbf{Check Blob Listeners.}
This analysis directly stems from the UI command one since we manually checked whether the detected \bls are correct with the threshold value of three commands per UI listener.
We used the same metrics used for the UI command detection to measure the accuracy of the \bls detection:

\begin{equation}\textstyle\label{eq.recallBlob}\small
recall_{blob} (\%) = \frac{|\{ correctBlobs\} \cap \{ detectedBlobs\}|}{|\{ correctBlobs\}|} \times 100
\end{equation}

\begin{equation}\textstyle\label{eq.precBlob}\small
precision_{blob} (\%) = \frac{|\{ correctBlobs\} \cap \{ detectedBlobs\}|}{|\{ detectedBlobs\}|} \times 100
\end{equation}

\subsection{Results and Analysis}\label{sub.results}

\noindent\textbf{RQ4: Command Detection Accuracy.} 
\Cref{tab.evalCmd} shows the number of commands successfully detected per software system.
\tool detected 1392 of the 1400 UI commands (eight false negatives), leading to a recall of \SI{99.43}{\percent}.
\tool also detected 62 irrelevant commands, leading to a precision of \SI{95.73}{\percent}.

\begin{table}[h]\small
 \caption{Command detection results}\label{tab.evalCmd}
  \centering \setlength{\tabcolsep}{2.9pt}
    \begin{tabular}{lSSSSS} 
    \toprule
	\textbf{Software} & \textbf{Successfully} 			&\textbf{FN} 	&\textbf{FP} 	&$\mathbf{Recall_{cmd}}$& $\mathbf{Precision_{cmd}}$\\
		\textbf{System}	   & \textbf{Detected}&\textbf{(\#)}	&\textbf{(\#)} &\textbf{(\%)}			    &\textbf{(\%)}\\	
	                  	& \textbf{Commands (\#)}     & & &  & \\	
 	\cmidrule(lr){1-1} \cmidrule(lr){2-2} \cmidrule(lr){3-3} \cmidrule(lr){4-4}  \cmidrule(lr){5-5} \cmidrule(lr){6-6}
  	Eclipse & 330 & 0 & 5 & 100 & 98.51 \\
  	JabRef & 510 & 5 & 7 & 99.03 & 98.65 \\
  	ArgoUML & 264 & 3 & 3 & 98.88 & 98.88\\
  	FreeCol & 288 & 0 & 47 & 100 & 85.93\\
	\cmidrule(lr){1-6}
	\textbf{OVERALL} & 1392 & 8 & 62 & 99.43 & 95.73\\
	\bottomrule
   \end{tabular}
\end{table}

\begin{figure}[h]
	\centering
		\includegraphics[width=0.99\columnwidth]{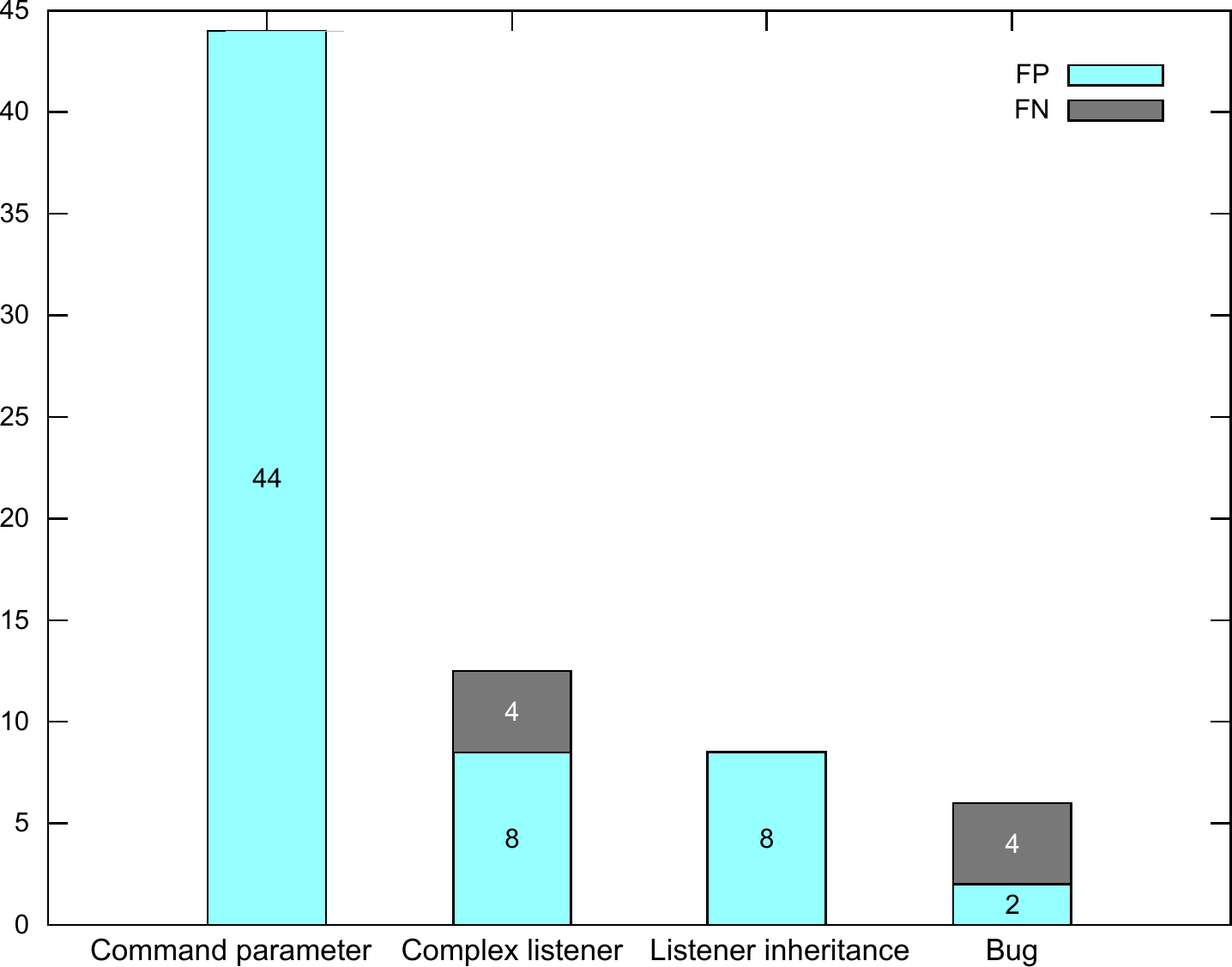}
		\caption{Distribution of the false negative and positive commands}\label{fig.fnCmds}
\end{figure}

\Cref{fig.fnCmds} classifies the 70 FN and FP commands according to their underlying issues.
44 FP commands are classified as \emph{command parameters}.
In these cases, several commands spotted in a listener are in fact a single command that is parametrized differently following the source interactive object.
The following code excerpt, from \emph{ArgoUML}, illustrates this case.
Two commands are detected by our algorithm (\Cref{param1,param2}).
These two commands seem to form a single command that takes a parameter, here the string parameter of \emph{setText}.

\begin{lstlisting}[language=MyJava, escapechar=~]
public void stateChanged(ChangeEvent ce) {
  JSlider srcSlider = (JSlider) ce.getSource();
  Goal d = (Goal) slidersToDecisions.get(srcSlider);
  JLabel valLab = (JLabel) slidersToDigits.get(srcSlider);
  int pri = srcSlider.getValue();
  d.setPriority(pri);
  
  if (pri == 0) {
    valLab.setText(Translator.localize("label.off"));~\label{param1}~
  } else {
    valLab.setText("    " + pri);~\label{param2}~
  }
 }
\end{lstlisting}

We marked 12 commands are being part of \emph{complex listeners}.
A complex listener involves complex and deeply nested conditional statements so that the algorithm fails to detect the commands properly.
This category could be part of the bug category since it is a limit of \tool but we wanted to show the limit of the proposal on complex conditional statements.

Eight listeners have been put in the \emph{listener inheritance} category.
This case refers to interactive objects checked several times in the listener inheritance hierarchy.
The following code excerpt illustrates this case.
Two commands (\Cref{dup2,dup1}) are detected by our algorithm for the UI listener of \emph{Panel1}.
In fact, these two commands refer to the same interactive object \emph{ok} and the execution of the first command (\Cref{dup2}) prevents the execution of the second one (\Cref{dup1}).
We consider this case as a design issue of the code.

\begin{lstlisting}[language=MyJava, escapechar=~]
public class Panel0 implements ActionListener {
  Button ok;
  
  public Panel0() {
    ok = new JButton();
    ok.setActionCommand("OK");
  }

  public void actionPerformed(ActionEvent ae) {
    if(ae.getActionCommand().equals("OK")) {~\label{dup1}~
      //...
    }
    //...
  }
}

public class Panel1 extends Panel0 {
  public void actionPerformed(ActionEvent ae) {
    if(ae.getActionCommand().equals("OK")) {~\label{dup2}~
     //...
     return;
    }
    super.actionPerformed(ae);
  }
}
\end{lstlisting}

Finally, the \emph{bug} category refers to various errors in \tool.

To conclude on RQ4, our approach is efficient for detecting UI commands that compose UI listener, even if improvements still possible.

\medskip
\noindent\textbf{RQ5: Blob Listeners Detection Accuracy.}
To validate that the refactoring is behavior-preserving, the refactored software systems have been manually tested by their developers we contacted and ourselves.
Test suites of each system have also been used.
\Cref{tab.evalBlobs} gives an overview of the results of the \bls detection per software system.
12 false positives and one false negative have been identified against 52 \bls correctly detected.
The average recall is \SI{98.11}{\percent} and the average precision is \SI{81.25}{\percent}.
The average time (computed on five executions for each software system) spent to analyze the software systems is \SI{5.9}{\second}.
It excludes the time that \emph{Spoon} takes to load all the classes, that is an average of \SI{22.4}{\second} per software system.
We did not consider the time that \emph{Spoon} takes since it is independent of our approach.

\begin{table}[h]\small
 \caption{\bl detection results}\label{tab.evalBlobs}
  \centering \setlength{\tabcolsep}{1.2pt}
    \begin{tabular}{lSSSSSr} 
    \toprule
	\textbf{Software} & \textbf{Successfully} 	  &\textbf{FN}  &\textbf{FP}    &$\mathbf{Recall_{blob}}$&$\mathbf{Precision_{blob}}$ &\textbf{Time} \\
  	\textbf{System}   & \textbf{Detected}	         &\textbf{(\#)}&\textbf{(\#)}  &\textbf{(\%)}		      &\textbf{(\%)}			         &\textbf{(ms)}\\
  	& \textbf{\bls (\#)} & & & & & \\	
 	\cmidrule(lr){1-1} \cmidrule(lr){2-2} \cmidrule(lr){3-3} \cmidrule(lr){4-4}  \cmidrule(lr){5-5} \cmidrule(lr){6-6}\cmidrule(lr){7-7}
  	Eclipse & 16  & 0  & 2  & 100  & 88.89 & 4 \\
  	JabRef  & 8   & 0  & 3  & 100  & 72.73  & 5.6\\
  	ArgoUML & 13  & 1  & 2  & 92.86  & 86.7  & 8.6\\
  	FreeCol & 15  & 0  & 5  & 100  & 75  & 5.5\\
	\cmidrule(lr){1-7}
	\textbf{OVERALL} & 52  & 1  & 12  & 98.11  & 81.25  & 5.9\\
	\bottomrule
   \end{tabular}
\end{table}

The FP and FN \bls is directly linked to the FP and FN of the commands detection.
For example, FP commands increased the number of commands in their listener to two or more so that such a listener is wrongly considered as a \bl.
This is the case for \emph{FreeCol} where 47 FP commands led to 5 FP \bls.

To conclude on RQ5, regarding the recall and the precision, our approach is efficient for detecting \bls.

\medskip
\noindent\textbf{RQ6: Blob Listeners Refactoring Accuracy.}

\begin{table}[h]\small
 \caption{\bl refactoring results}\label{tab.refactBlobs}
  \centering \setlength{\tabcolsep}{4pt}
    \begin{tabular}{lSSSr} 
    \toprule
	\textbf{Software} & \textbf{Successfully} 	  &\textbf{Failures}  &$\mathbf{Precision_{refact}}$  &\textbf{Time} \\
  	\textbf{System}   & \textbf{Refactored}	         &\textbf{(\#)}& \textbf{(\%)}  &\textbf{(s)}\\
  	& \textbf{\bls (\#)} & & & \\	
 	\cmidrule(lr){1-1} \cmidrule(lr){2-2} \cmidrule(lr){3-3} \cmidrule(lr){4-4}  \cmidrule(lr){5-5}
  	Eclipse & 4    & 12  & 25     & 133 \\
  	JabRef  & 4    & 4   & 50     & 236 \\
  	ArgoUML & 11   & 3   & 78.57  & 116  \\
  	FreeCol & 7    & 8   & 46.7   & 135 \\
	\cmidrule(lr){1-5}
	\textbf{OVERALL} & 26   & 27   & 49.06   & 155 \\
	\bottomrule
   \end{tabular}
\end{table}

This research question aims to provide quantitative results regarding the refactoring of \bls.
The results of \tool on the four software systems are described in \Cref{tab.refactBlobs}.
The average refactoring time (\ie five executions for each software system) is \SI{155}{\second}.
Most of the time is spent in find interactive objects in the code and their usages.
We think that optimizations can be done to improve this average time.
The average rate of \bls successfully refactored using the technique proposed in \Cref{sec.refactor} is \SI{55.1}{\percent}.
27 of the 49 \bls have been refactored.
Two main reasons explain this result:
1/ There exists in fact two types of \bls and our refactoring solution supports one of them;
2/ The second type of \bls may not be refactorable because of limitations of the Java GUI toolkits.
\Cref{tab.listRefact} details the results according to the type of the \bls.
\SI{86.7}{\percent} of the \bls that manage several listeners can be refactored.
The four failures of this type are related to bugs in the process.

\begin{table}[h]\small
 \caption{Refactoring results considering the type of the \bls}\label{tab.listRefact}
  \centering \setlength{\tabcolsep}{2pt}
    \begin{tabular}{lSS} 
    \toprule
	\textbf{Listener Type} & \textbf{Successes (\#)} 	  &\textbf{Failures (\#)}\\
 	\cmidrule(lr){1-1} \cmidrule(lr){2-2} \cmidrule(lr){3-3}
 	Several interactive objects for several & 26 & 4 \\
 	commands (\eg action listeners) &  &  \\
 	\cmidrule(lr){1-1} \cmidrule(lr){2-2} \cmidrule(lr){3-3}
  	One interactive object for several & 0 & 23 \\
  	commands (\eg mouse and key listeners) &   &  \\
	\bottomrule
   \end{tabular}
\end{table}

\begin{table*}[h]
 \caption{Commits and discussions}\label{tab.commits} \small
  \centering
    \begin{tabular}{ll}
    \toprule
\textbf{Software System}&   \textbf{Bug reports, commits, and discussions}\\
 	\cmidrule(lr){1-1}\cmidrule(lr){2-2}
Eclipse (platform.ui) & \scriptsize\url{https://bugs.eclipse.org/bugs/show_bug.cgi?id=510745}\\
& \scriptsize\url{https://dev.eclipse.org/mhonarc/lists/platform-ui-dev/msg07651.html}\smallskip\\
JabRef & \scriptsize\url{https://github.com/JabRef/jabref/pull/2369/}\smallskip\\
& \scriptsize\url{https://github.com/JabRef/jabref/commit/021f094e64a6393a7490ee680d73ef26b3128625}\\
ArgoUML & \scriptsize \url{http://argouml.tigris.org/issues/show_bug.cgi?id=6524}\smallskip\\
FreeCol &\scriptsize \url{https://sourceforge.net/p/freecol/mailman/message/35566820/}\\
&\scriptsize \url{https://sourceforge.net/p/freecol/git/ci/669cf9c74b208c141cea27ee254292b3422d5718/}\\
&\scriptsize \url{https://sourceforge.net/p/freecol/git/ci/2865215d3712a8d4d4bd958c1b397c90460192da/}\\
&\scriptsize \url{https://sourceforge.net/p/freecol/git/ci/cdc689c7ae4bbac9fcc729477d5cc7e40ac4a90b/}\\
&\scriptsize \url{https://sourceforge.net/p/freecol/git/ci/0eedd71b269b6cf20ec00f0fc5a7da932ceaab4f/}\\
&\scriptsize \url{https://sourceforge.net/p/freecol/git/ci/973422623b52289481f328b27f12543a4b383f38/}\\
&\scriptsize \url{https://sourceforge.net/p/freecol/git/ci/985adc4de11ccd33648e99294e5d91319cb23aa0/}\\
&\scriptsize \url{https://sourceforge.net/p/freecol/git/ci/4fe44e747cb30a161d8657750aa75b6c57ea30ab/}\\
	\bottomrule
    \end{tabular}
\end{table*}

A \bl is a listener that can produce several UI commands.
The proposed refactoring process works when several interactive objects register the same listener to produce several commands (one command per interactive object).
However, in several cases a single interactive object registers a listener to produce several commands.
The following code excerpt, simplified from \emph{Jabref}, illustrates this case.
A single interactive object registers this listener that treats keyboard events.
Several commands, however, are produced in this listener (\Cref{key1,key2}).
Our refactoring solution cannot be applied on such listeners.

\begin{lstlisting}[language=MyJava, escapechar=~]
public void keyPressed(KeyEvent e) {
  //...
  if (e.isControlDown()) {
    switch (e.getKeyCode()) {
      case KeyEvent.VK_UP:
        frame.getGroupSelector().moveNodeUp(node);~\label{key1}~
        break;
      case KeyEvent.VK_DOWN:
        frame.getGroupSelector().moveNodeDown(node);~\label{key2}~
        break;
      //...
}
\end{lstlisting}

Refactoring solutions for such key listeners may exist for several GUI toolkits that support key bindings.
For example, in the following code we manually fragmented the initial listener two atomic Java Swing key bindings (\Cref{keybind1,keybind2}).
Such a refactoring strongly depends on the targeted UI toolkit.

\begin{lstlisting}[language=MyJava, escapechar=~]
InputMap im = textfield.getInputMap();
ActionMap a = textfield.getActionMap();

im.put(KeyStroke.getKeyStroke(KeyEvent.VK_UP, ~\label{keybind1}~
       InputEvent.CTRL_MASK), "up");
a.put("up", e -> frame.getGroupSelector().moveNodeUp(node));

im.put(KeyStroke.getKeyStroke(KeyEvent.VK_DOWN, ~\label{keybind2}~
       InputEvent.CTRL_MASK), "down");
a.put("down",e->frame.getGroupSelector().moveNodeDown(node));
\end{lstlisting}

Another example of listeners that cannot be refactored is depicted in the following code excerpt.
This mouse listener contains three commands (\Cref{mouselist1,mouselist2,mouselist3}).
Similarly than for the previous key listener, our refactoring solution cannot be applied on such listeners.
Moreover, to the best of our knowledge no GUI toolkit permits the definition of mouse bindings as in the previous code example.
Such a \bl cannot thus be refactored.

\begin{lstlisting}[language=MyJava, escapechar=~]
public void mousePressed(MouseEvent e) {
  if (e.getClickCount() == 1) {
    // ...~\label{mouselist1}~
  }else if (e.getButton() == MouseEvent.BUTTON3) {
    //...~\label{mouselist2}~
  } else if (e.getClickCount() > 1) {
    //...~\label{mouselist3}~
  }
}
\end{lstlisting}

To conclude on RQ6, the refactoring solution we propose is efficient for one of the two types of \bls.
Refactoring the second type of \bls may not be possible and depends on the targeted GUI toolkit.

\medskip
\noindent\textbf{RQ7: Relevance of the \bl.}

This last research question aims to provide qualitative results regarding the refactoring of \bls.
We computed code metrics before and after the refactoring to measure the impact of the changes on the code.
We also submitted patches that remove the found and refactorable \bls from the analyzed software systems.
We then asked their developers for feedback regarding the patches and the concept of \bl.
The bug reports that contain the patches, the commits that remove \bls, and the discussions are listed in \Cref{tab.commits}.
Once the refactored code automatically produced using \tool, we manually applied some changes to follow the coding conventions of the different software systems.
Then, the patches have been manually created by applying a \emph{diff} between the original code and the automatically refactored one.
The patches submitted to \emph{Jabref} and \emph{FreeCol} have accepted and merged.
The patches for Eclipse are not yet merged but were positively commented.
We did not receive any comment regarding the patches for \emph{ArgoUML}.
We noticed that \emph{ArgoUML} is no more actively maintained.

We asked developers whether they consider that coding UI listeners that manage several interactive objects is a bad coding practice.
The developers that responded globally agree that \bl is a design smell.
"\emph{It does not strictly violate the MVC pattern. [...] Overall, I like your solution}".
"\emph{Probably yes, it depends, and in examples you've patched this was definitely a mess}".
An Eclipse developer suggest to complete the Eclipse UI development documentation to add information related to UI design smells and \bl.

Regarding the relevance of the refactoring solution:
"\emph{I like it when the code for defining a UI element and the code for interacting with it are close together.
So hauling code out of the action listener routine and into a lambda next to the point a button is defined is an obvious win for me.}"
A developer, however, explained that 
"\emph{there might be situations where this can not be achieved fully, e.g. due to limiting implementations provided by the framework.}"
We agree with this statement and the outcomes of RQ6 explain the limitations.
"\emph{It depends, if you refactor it by introducing duplicated code, then this is not suitable and even worse as before}".
We also agree with this statement that we commented by computing several code metrics before and after the refactoring, as summarized in \Cref{table.metrics}.
We computed the changes in terms of number of lines of code (LoC), cyclomatic complexity (CC), and duplicated lines (DUP).
The refactoring of \bls reduces the number of lines of code (-210 LoCs) and the cyclomatic complexity (-150 CC).
We noticed 25 duplicated lines of code.
These duplicated lines are pre- and post-statements of commands (see \Cref{fig.cmd}), \eg variable declarations, used by several commands in the same listener.
These statements are thus duplicated when the listener is separated into atomic ones.

\begin{table}[h]\small
 \caption{Code metrics changes with \bls refactoring}\label{table.metrics}
  \centering \setlength{\tabcolsep}{3pt}
    \begin{tabular}{lSSS} 
    \toprule
	\textbf{Software System} & \textbf{LoC (\#)} 	  &\textbf{CC (\#)} & \textbf{DUP (\#)}\\
 	\cmidrule(lr){1-1} \cmidrule(lr){2-2} \cmidrule(lr){3-3} \cmidrule(lr){4-4}
 	Eclipse & -40 & -45 & 11 \\
 	JabRef & -49 & -21 & 0 \\
 	ArgoUML & -35 & -47 & 13\\
 	FreeCol & -146 & -37 & 1 \\
 	\midrule
 	\textbf{OVERALL} & -270 & -150 & 25 \\
	\bottomrule
   \end{tabular}
\end{table}

To conclude on RQ7, the concept of \bl and the refactoring solution we propose is accepted by the developers we interviewed.
The refactoring has a positive impact on the code quality.
The interviews did not permit the identification of how \bl are introduced in the code, following the classification of Sharma \etal~\cite{Sharma18}.

\subsection{Threats to validity} \label{sub.threats}

\textbf{External validity.} 
This threat concerns the possibility to generalize our findings.
We designed the experiments using multiple Java Swing and SWT open-source software systems to diversify the observations.
These unrelated software systems are developed by different persons and cover various user interactions.
Our implementation and our empirical study (\Cref{sec.study}) focus on the Java Swing and SWT toolkits.
We focused on the Java Swing and SWT toolkits because of their popularity and the large quantity of Java Swing and SWT legacy code.
We provide on the companion web page examples of \bls in other Java UI toolkits, namely GWT and JavaFX\footnoteref{foot.webpage}.

\textbf{Construct validity.}
This threat relates to the perceived overall validity of the experiments.
Regarding the empirical study (\Cref{sec.study}), we used \tool to find UI commands in the code.
\tool might not have detected all the UI commands.
We showed in the validation of this tool (\Cref{sec.eval}) that its precision (\num{95.7}) and recall (\num{99.49}) limit this threat.
Regarding the validation of \tool, the detection of FNs and FPs have required a manual analysis of all the UI listeners of the software systems.
To limit errors during this manual analysis, we added a debugging feature in \tool for highlighting UI listeners in the code.
We used this feature to browse all the UI listeners and identify their commands to state whether these listeners are \bls.
During our manual analysis, we did not notice any error in the UI listener detection.
We also manually determined whether a listener is a \bl.
To reduce this threat, we carefully inspected each UI command highlighted by our tool.

\subsection{Scope of the Approach}
Our approach has the following limitations.
First, the command detection algorithm can be improved by detecting command parameters, \ie commands in a same listener that form a single command that can be parametrized.
Second, the refactoring of key listeners, using key bindings, can be supported by \tool.
This support may vary and depend on the targeted UI toolkit.

The current implementation of the approach supports JavaFX, Swing, and SWT code.
The approach is, however, not limited to these graphical toolkits and Java.
The concept of UI listener (at the origin of the Blob Listener) is used, to the best of our knowledge, by most of the modern graphical libraries usually along with data binding.

\section{Related Work}\label{sec.related}

\begin{table*}[h]\small
 \caption{UI design smells proposed in related works}\label{tab.uismells} 
  \centering \setlength{\tabcolsep}{2.2pt}
    \begin{tabular}{lllll} 
    \toprule
\textbf{Name}& \textbf{Description} & \textbf{Focus}~\cite{Sharma18} & \textbf{Structural} & \textbf{Detection}\\
& & & \textbf{Principle}~\cite{ganesh2013} & \textbf{Strategy}~\cite{Sharma18} \\
\cmidrule(lr){1-1}\cmidrule(lr){2-2}\cmidrule(lr){3-3}\cmidrule(lr){4-4}\cmidrule(lr){5-5}
\textbf{Blob Listener} & \textbf{A UI listener/handler that controls too} & Implementation & Deficient encapsulation & Metric-based\\
& \textbf{much interactive objects} & &\\
\cmidrule(lr){1-1}\cmidrule(lr){2-2}\cmidrule(lr){3-3}\cmidrule(lr){4-4}\cmidrule(lr){5-5}
Promiscuous Controller~\cite{aniche2017} & web server-side controllers that & Design and & Weakened modularity& Metric-based\\
& manage too many routes &Implementation \\
\cmidrule(lr){1-1}\cmidrule(lr){2-2}\cmidrule(lr){3-3}\cmidrule(lr){4-4}\cmidrule(lr){5-5}
Brain Controller~\cite{aniche2017} & Server-side controllers that do too much & Design and & Weakened modularity& Metric-based\\
& (lack of separation of concern) & Implementation\\
\cmidrule(lr){1-1}\cmidrule(lr){2-2}\cmidrule(lr){3-3}\cmidrule(lr){4-4}\cmidrule(lr){5-5}
UI Shotgun Surgery~\cite{Almeida2015} & A change on the UI structure spans over & Usability & N/A & None\\
 & multiple UI implementations\\ 
\cmidrule(lr){1-1}\cmidrule(lr){2-2}\cmidrule(lr){3-3}\cmidrule(lr){4-4}\cmidrule(lr){5-5}
Too Many Layers~\cite{Almeida2015} & A UI is composed of too many & Usability & N/A & None\\
 & layers (\eg windows) \\
\cmidrule(lr){1-1}\cmidrule(lr){2-2}\cmidrule(lr){3-3}\cmidrule(lr){4-4}\cmidrule(lr){5-5}
UI Middle Man~\cite{Almeida2015} & A UI component (\eg a window) delegates & Usability & N/A & None\\
& the job to another UI component &\\
\cmidrule(lr){1-1}\cmidrule(lr){2-2}\cmidrule(lr){3-3}\cmidrule(lr){4-4}\cmidrule(lr){5-5}
Information Overload~\cite{Almeida2015} & Too much information is provided to users & Usability & N/A & None\\
\cmidrule(lr){1-1}\cmidrule(lr){2-2}\cmidrule(lr){3-3}\cmidrule(lr){4-4}\cmidrule(lr){5-5}
UI Inappropriate Intimacy~\cite{Almeida2015} & Several UIs, accessible from different places, & Usability & N/A & None\\
& handle related domain elements\\
\cmidrule(lr){1-1}\cmidrule(lr){2-2}\cmidrule(lr){3-3}\cmidrule(lr){4-4}\cmidrule(lr){5-5}
UI Feature Envy~\cite{Almeida2015} & One UI allows user to perform a task also & Usability & N/A & None\\
& provided by another UI of the system  &\\
	\bottomrule
    \end{tabular}
\end{table*}

Work related to this paper fall into three categories:
design smell detection;
UI maintenance and evolution;
UI testing.

\subsection{Design Smell Detection}

The characterization and detection of object-oriented (OO) design smells have been widely studied~\cite{Sharma18,rasool2015review}. 
For instance, research works characterized various OO design smells associated with code refactoring operations~\cite{fowler1999,brown1998}.
Multiple empirical studies have been conducted to observe the impact of several OO design smells on the code.
These studies show that OO design smells can have a negative impact on maintainability~\cite{Yamashita2013}, understandability~\cite{Abbes2011}, and change- or fault-proneness~\cite{Khomh2012}.
While developing seminal advances on OO design smells, these research works focus on OO concerns only.
Improving the validation and maintenance of UI code implies a research focus on UI design smells, as we propose in this paper.

Related to UI code analysis, \Cref{tab.uismells} summarizes several UI design smells discussed in the next paragraphs.
Surveys and classifications~\cite{Sharma18,rasool2015review} on OO design smells identified characteristics of code smells that can be applied to UI design smells.
We used several of these characteristics to describe the UI design smells of \Cref{tab.uismells}:
\emph{focus}: the focus area where a design smell operates~\cite{Sharma18};
\emph{structural principle}: what structural principles a design smell violates~\cite{ganesh2013};
\emph{detection strategy}: the techniques used to detect a design smell~\cite{Sharma18}.
The \bl focuses on the implementation of controllers.
Interviews with developers and code exhibited that a problem of encapsulation:
UI listeners may be visible as super interfaces or classes of controllers.
We use a metric and code analysis techniques to detect \bl instances.

Aniche~\etal define several design smells that affect web applications~\cite{aniche2017}.
In particular, a design smell focuses on web controllers that bind the model (on the server side) and the UI (on the client side) of the application.
They define six web design smells.
Four of them concern the model component and are thus not related to UI.
The two remaining smells concern web controllers, \ie the server-side objects that receives queries from the client side.
The \emph{Promiscuous Controller} smell refers to \emph{a [web] controller offering too many actions}.
The detection of such a controller is based on the number of web routes (10) and services (3) it contains.
The \emph{Brain Controller} smell refers to a lack of separation of concerns in a web controller so that it becomes too complex.
These two design smells refer to implementation and design issues, and a weakness in the modularity of web controllers.
Metrics are used to identified these design smells in the code.
Web controllers and UI controllers strongly differ (web routes and services \emph{vs} interactive objects and UI listeners) so that our code analyses cannot be compared.
However, their results regarding promiscuous controllers follow ours on \bls:
web or UI controllers should not do too much.

Silva \etal propose an approach to inspect UI source code as a reverse engineering process~\cite{silva2010gui,Silva2010}.
Their goal is to provide developers with a framework supporting the development of UI metrics and code analyses.
They also applied standard OO code metrics on UI code~\cite{silva2014approach}.
Closely, Almeida \etal propose a set of UI smells that focus on usability~\cite{Almeida2015}.
These smells are described in \Cref{tab.uismells}.
Several of them (\emph{UI Shotgun Surgery}, \emph{UI Middle Man}, \emph{UI Inappropriate Intimacy}, and \emph{UI Feature Envy}) are adaptations of the object-oriented design smells of the same name~\cite{fowler1999} with a focus on UI code.
The \emph{Too Many Layers} and \emph{Information Overload} UI design smells are related to the structural complexity of UIs that may have a negative impact on their understanding by users.
The relevance of these UI smells are not studied by the authors.
These smells, however, show that UIs are software artifacts that require specific analysis techniques to improve their quality.

The automatic detection of design smells involves two steps.
First, a source code analysis is required to compute source code metrics. 
Second, heuristics are applied to detect design smells on the basis of the computed metrics to detect design smells.
Source code analyses can take various forms, notably: static, as we propose, and historical.
Regarding historical analysis, Palomba \etal propose an approach to detect design smells based on change history information~\cite{palomba14}.
Future work may also investigate whether analyzing code changes over time can help in characterizing \bls.
Regarding detection heuristics, the use of code metrics to define detection rules is a mainstream technique.
Metrics can be assemble with threshold values defined empirically to form detection rules~\cite{moha2010}. 
Search-based techniques are also used to exploit OO code metrics~\cite{Sahin14}, as well as machine learning~\cite{Zanoni2015}, or bayesian networks~\cite{khomh2011}.
Still, these works do not cover UI design smells.
In this paper, we focus on static code analysis to detect UI commands to form a \bl detection rule.
To do so, we used a Java source code analysis framework that permits the creation of specific code analyzers~\cite{spoon}.
Future work may investigate other heuristics and analyses to detect UI design smells.

Several research work on design smell characterization and detection are domain-specific.
For instance, Moha~\etal propose a characterization and a detection process of service-oriented architecture anti-patterns~\cite{moha12}.
Garcia \etal propose an approach for identifying architectural design smells~\cite{garcia2009}.
Similarly, this work aims at motivating that UIs form another domain concerned by specific design smells that have to be characterized.

Research studies have been conducted to evaluate the impact of design smells on system's quality~\cite{olbrich2010,fontana2013} or how they are perceived by developers~\cite{palomba2014}.
Future work may focus on how software developers perceive \bls.

\subsection{UI maintenance and evolution}

Unlike OO design smells, less research work focuses on UI design smells.
Zhang \etal{} propose a technique to automatically repair broken workflows in Swing UIs~\cite{ZhangLE2013}.
Static analyses are proposed.
This work highlights the difficulty "\emph{for a static analysis to distinguish UI actions [UI commands] that share the same event handler [UI listener]}".
In our work, we propose an approach to accurately detect UI commands that compose UI listeners.
Staiger also proposes a static analysis to extract UI code, interactive objects, and their hierarchies in C/C++ software systems~\cite{Staiger07}.
The approach, however, is limited to find relationships between UI elements and thus does not analyze UI controllers and their listeners.
Zhang \etal{} propose a static analysis to find violations in UIs~\cite{Zhang2012}.
These violations occur when UI operations are invoked by non-UI threads leading a UI error.
The static analysis is applied to infer a static call graph and check the violations.
Frolin \etal{} propose an approach to automatically find inconsistencies in MVC JavaScript applications~\cite{Frolin15}.
UI controllers are statically analyzed to identify consistency issues (\eg inconsistencies between variables and controller functions).
This work is highly motivated by the weakly-typed nature of Javascript.

\subsection{UI testing}

UI testing may also implies UI code analysis techniques.
The automatic production of UI tests requires techniques to extract UI information at design or run time.
Such techniques may involve the dynamic detection of interactive objects to automatically interact with them at run time~\cite{nguyen2014}.
Several UI testing techniques improve the automatic interaction with UIs by analyzing in the code the dependencies between the interactive objects and the UI listeners~\cite{arlt2012,yang2015,yang2015static,ganov2009event}.
UI listeners are analyzed to:
identify class fields shared by several listeners~\cite{arlt2012};
detect dependencies between UI listeners~\cite{yang2015};
detect transitions between graphical windows~\cite{yang2015static};
identify, from one-command listeners, relevant input data to test interactive objects and produce UI tests~\cite{ganov2009event}.

\section{Conclusion and Future Work}\label{sec.conclu}

\subsection{Conclusion}

In this paper, we investigated a new research area on UI design smells.
We detailed a specific UI design smell, we called \bl, that can affect UI listeners.
The empirical study we conducted exhibits a specific number of UI commands per UI listener that characterizes a \bl exists.
We defined this threshold to three UI commands per UI listener.
We detailed an algorithm to automatically detect \bls.
We then proposed a behavior-preserving algorithm to refactor detected \bls.
The detection and refactoring algorithms have been implemented in a tool publicly available and then evaluated.
The results showed that the \bls detection and refactoring techniques are effective with several possible improvements.
Developers agreed that the \bls is a design smell.

\subsection{Research Agenda}

UI code is a major part of the code of software systems more and more interactive.
We argue that code analysis tools, similar to \emph{Findbugs} or \emph{PMD} that analyze object-oriented code, should be developed to specifically detect UI design smells.
\tool is a first step in this way.
In our future work, we first aim to complete \tool with other UI design smells that we would empirically identify and characterize.
We will investigate whether some UI faults~\cite{LEL15} are accentuated by UI design smells.
Second, the current UI event handling, based on the observer pattern, faces several critical limits~\cite{maier2012deprecating}.
UI event handling, however, is still used by a majority of modern UI libraries along with data binding mechanisms.
We think that UI event handling should be replaced by reactive programming / data binding approaches specifically designed to react on UI events.

\section*{Acknowledgements}
We thank Yann-Ga\"el Gu\'eh\'eneuc for his insightful comments on this paper.
We also thank Jean-R\'emy Falleri for his help on the statistical analysis.

\section*{References}
\bibliographystyle{elsarticle-num}
\bibliography{ref}

\begin{thebibliography}{10}
\expandafter\ifx\csname url\endcsname\relax
  \def\url#1{\texttt{#1}}\fi
\expandafter\ifx\csname urlprefix\endcsname\relax\def\urlprefix{URL }\fi
\expandafter\ifx\csname href\endcsname\relax
  \def\href#1#2{#2} \def\path#1{#1}\fi

\bibitem{Krasner88}
G.~E. Krasner, S.~T. Pope, A description of the {Model-View-Controller} user
  interface paradigm in {Smalltalk80} system, Journal of Object Oriented
  Programming 1 (1988) 26--49.

\bibitem{potel1996}
M.~Potel, {MVP: Model-View-Presenter the Taligent programming model for C++ and
  Java}, Taligent Inc.

\bibitem{Myers91}
B.~A. Myers, Separating application code from toolkits: Eliminating the
  spaghetti of call-backs, in: Proceedings of the 4th Annual ACM Symposium on
  User Interface Software and Technology, UIST '91, 1991, pp. 211--220.
\newblock \href {http://dx.doi.org/10.1145/120782.120805}
  {\path{doi:10.1145/120782.120805}}.

\bibitem{smith09}
J.~Smith, \href{http://msdn.microsoft.com/en-us/magazine/dd419663.aspx}{{WPF}
  apps with the model-view-viewmodel design pattern}, MSDN Magazine.
\newline\urlprefix\url{http://msdn.microsoft.com/en-us/magazine/dd419663.aspx}

\bibitem{palomba14}
F.~Palomba, G.~Bavota, M.~Di~Penta, R.~Oliveto, D.~Poshyvanyk, A.~De~Lucia,
  Mining version histories for detecting code smells, Software Engineering,
  IEEE Transactions on\href {http://dx.doi.org/10.1109/TSE.2014.2372760}
  {\path{doi:10.1109/TSE.2014.2372760}}.

\bibitem{Khomh2012}
F.~Khomh, M.~D. Penta, Y.-G. Gu\'{e}h\'{e}neuc, G.~Antoniol, {An exploratory
  study of the impact of antipatterns on class change- and fault-proneness},
  Empirical Software Engineering 17~(3) (2012) 243--275.
\newblock \href {http://dx.doi.org/10.1007/s10664-011-9171-y}
  {\path{doi:10.1007/s10664-011-9171-y}}.

\bibitem{lozano2007assessing}
A.~Lozano, M.~Wermelinger, B.~Nuseibeh, Assessing the impact of bad smells
  using historical information, in: Workshop on Principles of software
  evolution, ACM, 2007, pp. 31--34.
\newblock \href {http://dx.doi.org/10.1145/1294948.1294957}
  {\path{doi:10.1145/1294948.1294957}}.

\bibitem{rapu2004using}
D.~Rapu, S.~Ducasse, T.~G{\^\i}rba, R.~Marinescu, Using history information to
  improve design flaws detection, in: Proc. of Conference on Software
  Maintenance and Reengineering, 2004, pp. 223--232.
\newblock \href {http://dx.doi.org/10.1109/CSMR.2004.1281423}
  {\path{doi:10.1109/CSMR.2004.1281423}}.

\bibitem{aniche2017}
M.~Aniche, G.~Bavota, C.~Treude, M.~A. Gerosa, A.~van Deursen, Code smells for
  model-view-controller architectures, Empirical Software Engineering (2017)
  1--37\href {http://dx.doi.org/10.1007/s10664-017-9540-2}
  {\path{doi:10.1007/s10664-017-9540-2}}.

\bibitem{LEL16}
V.~Lelli, A.~Blouin, B.~Baudry, F.~Coulon, O.~Beaudoux,
  \href{https://hal.inria.fr/hal-01308625}{Automatic detection of {GUI} design
  smells: The case of blob listener}, in: {EICS'16: Proceedings of the 8th ACM
  SIGCHI symposium on Engineering interactive computing systems (EICS 2016)},
  ACM, 2016.
\newline\urlprefix\url{https://hal.inria.fr/hal-01308625}

\bibitem{fowler1999}
M.~Fowler, Refactoring: Improving the Design of Existing Code, Addison-Wesley,
  Boston, 1999.

\bibitem{W3CMBUI}
W3C, \href{https://www.w3.org/TR/abstract-ui/}{{MBUI} - abstract user interface
  models}, Tech. rep. (2014).
\newline\urlprefix\url{https://www.w3.org/TR/abstract-ui/}

\bibitem{GAM95}
E.~Gamma, R.~Helm, R.~Johnson, J.~Vlissides, Design patterns: elements of
  reusable object-oriented software, Addison-Wesley, 1995.

\bibitem{BEA00b}
M.~Beaudouin-Lafon, \href{10.1145/332040.332473}{Instrumental interaction: an
  interaction model for designing post-{WIMP} user interfaces}, in: Proc. of
  CHI'00, ACM, 2000, pp. 446--453.
\newline\urlprefix\url{10.1145/332040.332473}

\bibitem{BLO10}
A.~Blouin, O.~Beaudoux, \href{https://hal.inria.fr/inria-00477627}{Improving
  modularity and usability of interactive systems with {Malai}}, in: Proc. of
  EICS'10, 2010, pp. 115--124.
\newline\urlprefix\url{https://hal.inria.fr/inria-00477627}

\bibitem{BLO11}
A.~Blouin, B.~Morin, G.~Nain, O.~Beaudoux, P.~Albers, J.-M. J\'ez\'equel,
  Combining aspect-oriented modeling with property-based reasoning to improve
  user interface adaptation, in: EICS'11: Proceedings of the 3rd ACM SIGCHI
  symposium on Engineering interactive computing systems, 2011, pp. 85--94.
\newblock \href {http://dx.doi.org/10.1145/1996461.1996500}
  {\path{doi:10.1145/1996461.1996500}}.

\bibitem{XU05}
B.~Xu, J.~Qian, X.~Zhang, Z.~Wu, L.~Chen, A brief survey of program slicing,
  SIGSOFT Softw. Eng. Notes 30 (2005) 1--36.
\newblock \href {http://dx.doi.org/10.1145/1050849.1050865}
  {\path{doi:10.1145/1050849.1050865}}.

\bibitem{Frolin15}
F.~Ocariza, K.~Pattabiraman, A.~Mesbah, Detecting inconsistencies in
  {JavaScript} {MVC} applications, in: Proceedings of the ACM/IEEE
  International Conference on Software Engineering (ICSE), ACM, 2015, p. 11
  pages.
\newblock \href {http://dx.doi.org/10.1109/ICSE.2015.52}
  {\path{doi:10.1109/ICSE.2015.52}}.

\bibitem{spoon}
R.~Pawlak, M.~Monperrus, N.~Petitprez, C.~Noguera, L.~Seinturier, {SPOON: A
  library for implementing analyses and transformations of Java source code},
  Software: Practice and Experience 43~(4).
\newblock \href {http://dx.doi.org/10.1002/spe.2346}
  {\path{doi:10.1002/spe.2346}}.

\bibitem{Abbes2011}
M.~Abbes, F.~Khomh, Y.~G. Gu{\'{e}}h{\'{e}}neuc, G.~Antoniol, {An empirical
  study of the impact of two antipatterns, Blob and Spaghetti Code, on program
  comprehension}, in: Proceedings of the European Conference on Software
  Maintenance and Reengineering, 2011, pp. 181--190.
\newblock \href {http://dx.doi.org/10.1109/CSMR.2011.24}
  {\path{doi:10.1109/CSMR.2011.24}}.

\bibitem{Sheskin2007}
D.~J. Sheskin, {Handbook Of Parametric And Nonparametric Statistical
  Procedures, Fourth Edition}, {Chapman \& Hall/CRC}, 2007.

\bibitem{Johnson2013}
B.~Johnson, Y.~Song, E.~Murphy-Hill, R.~Bowdidge, {Why don't software
  developers use static analysis tools to find bugs?}, 2013, pp. 672--681.
\newblock \href {http://dx.doi.org/10.1109/ICSE.2013.6606613}
  {\path{doi:10.1109/ICSE.2013.6606613}}.

\bibitem{palomba2014}
F.~Palomba, G.~Bavota, M.~D. Penta, R.~Oliveto, A.~D. Lucia, Do they really
  smell bad? a study on developers' perception of bad code smells, in: Proc. of
  ICSM'14, IEEE, 2014, pp. 101--110.
\newblock \href {http://dx.doi.org/10.1109/ICSME.2014.32}
  {\path{doi:10.1109/ICSME.2014.32}}.

\bibitem{Mens2004}
T.~Mens, T.~Tourw{\'{e}}, {A Survey of Software Refactoring}, IEEE Transactions
  on Software Engineering 30~(2) (2004) 126--139.
\newblock \href {http://dx.doi.org/10.1109/TSE.2004.1265817}
  {\path{doi:10.1109/TSE.2004.1265817}}.

\bibitem{Sharma18}
T.~Sharma, D.~Spinellis, A survey on software smells, Journal of Systems and
  Software 138 (2018) 158 -- 173.
\newblock \href {http://dx.doi.org/10.1016/j.jss.2017.12.034}
  {\path{doi:10.1016/j.jss.2017.12.034}}.

\bibitem{ganesh2013}
S.~Ganesh, T.~Sharma, G.~Suryanarayana,
  \href{http://www.jot.fm/issues/issue_2013_06/article1.pdf}{Towards a
  principle-based classification of structural design smells.}, Journal of
  Object Technology 12~(2) (2013) 1--1.
\newline\urlprefix\url{http://www.jot.fm/issues/issue_2013_06/article1.pdf}

\bibitem{Almeida2015}
D.~Almeida, J.~C. Campos, J.~a. Saraiva, J.~a.~C. Silva, Towards a catalog of
  usability smells, in: Proceedings of the 30th Annual ACM Symposium on Applied
  Computing, SAC '15, ACM, 2015, pp. 175--181.
\newblock \href {http://dx.doi.org/10.1145/2695664.2695670}
  {\path{doi:10.1145/2695664.2695670}}.

\bibitem{rasool2015review}
G.~Rasool, Z.~Arshad, {A review of code smell mining techniques}, Journal of
  Software: Evolution and Process 27~(11) (2015) 867--895.
\newblock \href {http://dx.doi.org/10.1002/smr.1737}
  {\path{doi:10.1002/smr.1737}}.

\bibitem{brown1998}
W.~J. Brown, H.~W. McCormick, T.~J. Mowbray, R.~C. Malveau, AntiPatterns:
  refactoring software, architectures, and projects in crisis, Wiley New York,
  1998.

\bibitem{Yamashita2013}
A.~Yamashita, L.~Moonen, {Exploring the impact of inter-smell relations on
  software maintainability: an empirical study}, in: 35th International
  Conference on Software Engineering, ICSE '13, 2013, pp. 682--691.
\newblock \href {http://dx.doi.org/10.1109/ICSE.2013.6606614}
  {\path{doi:10.1109/ICSE.2013.6606614}}.

\bibitem{silva2010gui}
J.~C. Silva, J.~C. Campos, J.~A. Saraiva,
  \href{http://hdl.handle.net/1822/18517}{{GUI} inspection from source code
  analysis}.
\newline\urlprefix\url{http://hdl.handle.net/1822/18517}

\bibitem{Silva2010}
J.~a.~C. Silva, C.~Silva, R.~D. Gon\c{c}alo, J.~a. Saraiva, J.~C. Campos, The
  {GUISurfer} tool: Towards a language independent approach to reverse
  engineering {GUI} code, in: Proceedings of the 2Nd ACM SIGCHI Symposium on
  Engineering Interactive Computing Systems, EICS '10, ACM, 2010, pp. 181--186.
\newblock \href {http://dx.doi.org/10.1145/1822018.1822045}
  {\path{doi:10.1145/1822018.1822045}}.

\bibitem{silva2014approach}
J.~C. Silva, J.~C. Campos, J.~Saraiva, J.~L. Silva, An approach for graphical
  user interface external bad smells detection, in: New Perspectives in
  Information Systems and Technologies, 2014, pp. 199--205.
\newblock \href {http://dx.doi.org/10.1007/978-3-319-05948-8_19}
  {\path{doi:10.1007/978-3-319-05948-8_19}}.

\bibitem{moha2010}
N.~Moha, Y.-G. Gueheneuc, L.~Duchien, A.~Le~Meur, {DECOR}: A method for the
  specification and detection of code and design smells, Software Engineering,
  IEEE Transactions on 36~(1) (2010) 20--36.
\newblock \href {http://dx.doi.org/10.1109/TSE.2009.50}
  {\path{doi:10.1109/TSE.2009.50}}.

\bibitem{Sahin14}
D.~Sahin, M.~Kessentini, S.~Bechikh, K.~Deb, Code-smell detection as a bilevel
  problem, ACM Trans. Softw. Eng. Methodol. 24~(1) (2014) 6:1--6:44.
\newblock \href {http://dx.doi.org/10.1145/2675067}
  {\path{doi:10.1145/2675067}}.

\bibitem{Zanoni2015}
M.~Zanoni, F.~A. Fontana, F.~Stella, On applying machine learning techniques
  for design pattern detection, Journal of Systems and Software 103~(0) (2015)
  102 -- 117.
\newblock \href {http://dx.doi.org/10.1016/j.jss.2015.01.037}
  {\path{doi:10.1016/j.jss.2015.01.037}}.

\bibitem{khomh2011}
F.~Khomh, S.~Vaucher, Y.-G. Gu{\'e}h{\'e}neuc, H.~Sahraoui, {BDTEX}: A
  {GQM}-based bayesian approach for the detection of antipatterns, Journal of
  Systems and Software 84~(4) (2011) 559--572.
\newblock \href {http://dx.doi.org/doi.org/10.1016/j.jss.2010.11.921}
  {\path{doi:doi.org/10.1016/j.jss.2010.11.921}}.

\bibitem{moha12}
N.~Moha, F.~Palma, M.~Nayrolles, B.~Joyen~Conseil, G.~Yann-Gael, B.~Baudry,
  J.-M. J{\'e}z{\'e}quel,
  \href{https://hal.inria.fr/hal-00722472}{{Specification and Detection of SOA
  Antipatterns}}, in: {International Conference on Service Oriented Computing},
  2012.
\newline\urlprefix\url{https://hal.inria.fr/hal-00722472}

\bibitem{garcia2009}
J.~Garcia, D.~Popescu, G.~Edwards, N.~Medvidovic, Identifying architectural bad
  smells, in: Software Maintenance and Reengineering, 2009. CSMR'09. 13th
  European Conference on, IEEE, 2009, pp. 255--258.
\newblock \href {http://dx.doi.org/10.1109/CSMR.2009.59}
  {\path{doi:10.1109/CSMR.2009.59}}.

\bibitem{olbrich2010}
S.~M. Olbrich, D.~S. Cruzes, D.~I. Sjoberg, Are all code smells harmful? a
  study of god classes and brain classes in the evolution of three open source
  systems, in: 2010 IEEE International Conference on Software Maintenance,
  IEEE, 2010, pp. 1--10.
\newblock \href {http://dx.doi.org/10.1109/ICSM.2010.5609564}
  {\path{doi:10.1109/ICSM.2010.5609564}}.

\bibitem{fontana2013}
F.~A. Fontana, V.~Ferme, A.~Marino, B.~Walter, P.~Martenka, Investigating the
  impact of code smells on system's quality: An empirical study on systems of
  different application domains, in: Proc. of ICSM'13, IEEE, 2013, pp.
  260--269.
\newblock \href {http://dx.doi.org/10.1109/ICSM.2013.37}
  {\path{doi:10.1109/ICSM.2013.37}}.

\bibitem{ZhangLE2013}
S.~Zhang, H.~L{\"u}, M.~D. Ernst, Automatically repairing broken workflows for
  evolving {GUI} applications, in: ISSTA 2013, Proceedings of the 2013
  International on Software Testing and Analysis, 2013, pp. 45--55.
\newblock \href {http://dx.doi.org/10.1145/2483760.2483775}
  {\path{doi:10.1145/2483760.2483775}}.

\bibitem{Staiger07}
S.~Staiger, Static analysis of programs with graphical user interface, in:
  Software Maintenance and Reengineering, 2007. CSMR '07. 11th European
  Conference on, 2007, pp. 252--264.
\newblock \href {http://dx.doi.org/10.1109/CSMR.2007.44}
  {\path{doi:10.1109/CSMR.2007.44}}.

\bibitem{Zhang2012}
S.~Zhang, H.~L\"{u}, M.~D. Ernst, Finding errors in multithreaded {GUI}
  applications, in: Proceedings of the 2012 International Symposium on Software
  Testing and Analysis, ISSTA 2012, ACM, 2012, pp. 243--253.
\newblock \href {http://dx.doi.org/10.1145/2338965.2336782}
  {\path{doi:10.1145/2338965.2336782}}.

\bibitem{nguyen2014}
B.~N. Nguyen, B.~Robbins, I.~Banerjee, A.~Memon, {GUITAR}: an innovative tool
  for automated testing of {GUI}-driven software, Automated Software
  Engineering 21~(1) (2014) 65--105.
\newblock \href {http://dx.doi.org/10.1007/s10515-013-0128-9}
  {\path{doi:10.1007/s10515-013-0128-9}}.

\bibitem{arlt2012}
S.~Arlt, A.~Podelski, C.~Bertolini, M.~Sch{\"a}f, I.~Banerjee, A.~M. Memon,
  Lightweight static analysis for {GUI} testing, in: Software Reliability
  Engineering (ISSRE), 2012 IEEE 23rd International Symposium on, IEEE, 2012,
  pp. 301--310.
\newblock \href {http://dx.doi.org/10.1109/ISSRE.2012.25}
  {\path{doi:10.1109/ISSRE.2012.25}}.

\bibitem{yang2015}
S.~Yang, D.~Yan, H.~Wu, Y.~Wang, A.~Rountev, Static control-flow analysis of
  user-driven callbacks in android applications, in: Software Engineering
  (ICSE), 2015 IEEE/ACM 37th IEEE International Conference on, Vol.~1, IEEE,
  2015, pp. 89--99.
\newblock \href {http://dx.doi.org/10.1109/ICSE.2015.31}
  {\path{doi:10.1109/ICSE.2015.31}}.

\bibitem{yang2015static}
S.~Yang, H.~Zhang, H.~Wu, Y.~Wang, D.~Yan, A.~Rountev, Static window transition
  graphs for android (t), in: Automated Software Engineering (ASE), 2015 30th
  IEEE/ACM International Conference on, IEEE, 2015, pp. 658--668.
\newblock \href {http://dx.doi.org/10.1109/ASE.2015.76}
  {\path{doi:10.1109/ASE.2015.76}}.

\bibitem{ganov2009event}
S.~Ganov, C.~Killmar, S.~Khurshid, D.~E. Perry, Event listener analysis and
  symbolic execution for testing {GUI} applications, in: International
  Conference on Formal Engineering Methods, Springer, 2009, pp. 69--87.
\newblock \href {http://dx.doi.org/10.1007/978-3-642-10373-5_4}
  {\path{doi:10.1007/978-3-642-10373-5_4}}.

\bibitem{LEL15}
V.~Lelli, A.~Blouin, B.~Baudry,
  \href{https://hal.inria.fr/hal-01114724v1}{Classifying and qualifying {GUI}
  defects}, in: {IEEE International Conference on Software Testing,
  Verification and Validation (ICST 2015)}, IEEE, 2015.
\newline\urlprefix\url{https://hal.inria.fr/hal-01114724v1}

\bibitem{maier2012deprecating}
I.~Maier, M.~Odersky,
  \href{https://infoscience.epfl.ch/record/176887}{Deprecating the observer
  pattern with scala. react}, Tech. rep. (2012).
\newline\urlprefix\url{https://infoscience.epfl.ch/record/176887}

\end{thebibliography}

\end{document}